\documentclass[11pt,letter]{article}       % onecolumn (second format)

\usepackage{geometry}
\usepackage{graphicx}
\usepackage{amsmath}
\usepackage{amsmath,bm}
\usepackage{amsfonts}
\usepackage{amssymb}
\usepackage{graphicx}
\usepackage{subfigure}
\usepackage{float}
\usepackage{mathtools}
\usepackage{authblk}
\usepackage{cite}
\usepackage{etoolbox}
\usepackage{textcomp}
\apptocmd{\thebibliography}{\renewcommand{\sc}{}}{}{}
\usepackage{epstopdf}%

\title{Analysis and prediction of shock formation in acoustic energy transfer systems}

\author[1]{Vamsi C. Meesala}
\author[2]{Muhammad R. Hajj} 
\author[3]{Shima Shahab \thanks{Address all correspondence to this author. E-mail address: sshahab@vt.edu}}
\affil[1]{Department of Biomedical Engineering and Mechanics, Virginia Tech, 495 Old Turner Street, Norris Hall, Blacksburg, VA 24060, USA}
\affil[2]{Department of Civil, Environmental and Ocean Engineering, Davidson laboratory, Stevens Institute of Technology, 711 Hudson Street, Hoboken, NJ 07030, USA}
\affil[3]{Department of Mechanical Engineering, Virginia Tech, 635 Prices Fork Road, Goodwin Hall, Blacksburg, VA 24060, USA}
\date{}
\begin{document}
\maketitle
\vspace{-30pt}
\begin{abstract}
Losses associated with nonlinear wave propagation and exhibited by acoustic wave distortion and shock formation compromise the efficiency of contactless acoustic energy transfer systems. As such, predicting the shock formation distance and its dependence on the amplitude of the excitation is essential for their efficiency, design and operation. We present an analytical approach capable of predicting the shock formation distance of acoustic waves generated by a baffled disk with arbitrary deformation in a weakly viscous fluid medium. The loss-less Westervelt equation, used to model the nonlinear wave propagation, is asymptotically expanded based on the amplitude of the excitation. Because the solutions of the first- and second-order equations decay at different rates, we implement the method of renormalization and introduce a coordinate transformation to identify and eliminate the secular terms. The approach yields two partial differential equations that can be solved to predict the formation distance either analytically or numerically much faster than time-domain numerical simulations. The analysis and results are validated with solutions obtained from a nonlinear finite element simulation and previous experimental measurements.
\end{abstract}

\section{Introduction}
Acoustic energy transfer (AET) is a transformative contactless energy transfer (CET) technology that utilizes acoustic waves to transfer energy between piezoelectric transducers. AET systems have been shown to outperform conventional electromagnetic CET technologies \cite{villa2009design,bi2016review,maharjan2018indoor,jafari2019design,yu2011wireless,kim2012wireless} in critical applications \cite{roes2013acoustic,jang2019underwater}. In particular, AET has been proposed to recharge and communicate with low-power (e.g., 1$\mu$W-10mW) implanted medical devices  \cite{valdastri2011wireless,vaiarello2011ultra,maleki2011ultrasonically}, which eliminates the need for surgery to replace batteries \cite{lee2007first,maleki2011ultrasonically,charthad2015mm,seo2015model,charthad2018mm,jiang2019ultrasound,baltsavias2019vivo}; to develop battery-free underwater sensing networks to observe ocean conditions, track migration and habitats of marine animals, and  monitor oil spills \cite{leonard2010coordinated,domingo2012overview,xu2014applications,akyildiz2016softwater,jang2019underwater,li2020underwater}. These applications present the need to developing mathematical and numerical models capable of assessing the efficiency of AET systems \cite{ozeri2010ultrasonic,shahab2014contactless,shahab2015ultrasonic,gorostiaga2017analytic,gorostiaga2017experiment,bakhtiari2018acoustic,meesala2019modeling,basaeri2019mems,herrera2019aln,basaeri2019acoustic,allam2019aspect,meesala2020acoustic,bakhtiari2020dynamics,bhargava2020nonlinear,bhargava2020contactless}. In general, most of the current approaches neglect nonlinear effects associated with acoustic wave propagation \cite{hamilton1998nonlinear} and the electro-elastic response of the piezoelectric transmitters and receivers\cite{leadenham2015unified, meesala2018identification}. On the other hand, these effects become significant as the source strength is increased to enable higher energy transfer. As such, there is a need to expand the analysis capabilities to models that account for nonlinear effects and investigate their impact on the efficiency of energy transfer systems. In our previous work, we investigated theoretically the effect of material nonlinearity of a piezoelectric receiving disk on the energy transfer. We showed that the material nonlinearity can shift the optimum load resistance and that the shift is a function of the source strength \cite{meesala2019modeling}. We have also shown experimentally that the interplay of all the nonlinearities and the standing wave effects between the transmitter and receiver in an AET system can manifest themselves in a complex manner and have an impact on the energy transfer efficiency \cite{bhargava2020nonlinear}. 

One consequence of nonlinear acoustic wave propagation is exhibited by the distortion of its waveform due to a difference in the traveling speeds of the compression and rarefaction parts of the wave. In the frequency domain, this distortion is interpreted as energy transfer from the fundamental wave frequency to its higher harmonics. The accumulation of the distortion effect as the wave travels results in a discontinuity, referred to as a shock \cite{hamilton1998nonlinear}. The occurrence of a shock is associated with significant loss in energy that is proportional to the cube of the difference in pressure across the discontinuity. This loss compounds as the wave propagates further resulting in further reduction of the acoustic power \cite{rudnick1953attenuation}. It is relevant to point out here that in a weakly viscous fluid such as water and air, shocks occur before the attenuation is significant \cite{muir1980prediction}. In other words, attenuation effects can be neglected, and a lossless second-order wave equation can be used to analyze the nonlinear wave propagation. In the case of a finite amplitude plane wave, the amplitudes of the higher harmonics grow at the expense of the amplitude of the fundamental or excitation frequency up to the initial shock formation location $\overline{x}$. Beyond $\overline{x}$, all components decay due to the energy dissipated as a consequence of the shock propagation \cite{muir1980prediction}. In the context of AET,  the component of pressure at the excitation frequency $p_\omega$, that is generated by the transmitting disk operating under high excitation voltage decreases as the distance from the disk is increased due to diffraction and transfer of energy to higher harmonics. However, the total acoustic power remains conserved up to $\overline{x}$. Beyond $\overline{x}$, in addition to the transfer of energy and diffraction, the decrease in $p_\omega$ will be compounded by additional losses in energy due to the formation and propagation of shocks. In AET systems, the power transfer efficiency depends mostly on the pressure associated with the excitation frequency, $p_\omega$, as the higher harmonics do not necessarily coincide with the higher modes of the receiver \cite{bhargava2020contactless}. As such, the power transfer efficiency will be significantly compromised beyond $\overline{x}$, which renders this distance as an essential design parameter for high-intensity AET power transfer. 

Analytical expressions for $\overline{x}$ are readily available for plane waves, spherical waves, and on the axis of a focused Gaussian beam \cite{muir1980prediction,dalecki1991absorption}, which is not the case in AET system where the disk undergoes transverse deformations. Several numerical and analytical studies investigated nonlinear wave propagation and shock characteristics of acoustic wave generated by disks \cite{aanonsen1984distortion,lee1995time,khokhlova2001numerical,ginsberg1984nonlinear,ginsberg1984nonlinear2,coulouvrat1991analytical,fro1996renormalization,foda2014axial}. The numerical simulations provide an accurate description of the response. However, they are computationally expensive, especially when solved in the the time-domain. The objective of this effort is to develop an analytical approach to predict shock formation associated with a propagating acoustic wave generated by a vibrating disk with arbitrary transverse displacement. We consider an axisymmetric-baffled-vibrating piezoelectric disk and use the Westervelt equation to investigate the associated nonlinear wave propagation. In particular, we scale the governing equation and boundary conditions with $\epsilon$ and obtain analytical expressions for the $\epsilon-$order and $\epsilon^2-$order solutions using the Rayleigh integral. Next, we follow the work of Kelly and Nayfeh \cite{kelly1980non} with some modifications to eliminate the secular terms by implementing the method of renormalization \cite{nayfeh2008nonlinear,nayfeh2008perturbation} and obtain a uniformly valid solution of the Westervelt equation. We validate the predictions of the analytical approach with higher fidelity finite element simulations and previously published experimental results. 
%%%%%%%%%%%%%%%%%%%%%%%%%%%%%%%%%%%%%%%%%%%%%%%%%%%%%%%%%%%%%%%%%%%%%%%%%%%%%%%%%%%%%%%%%%%%%%%%%%%%%%%%%%%%%%%%%%%%%%%%%%%%%%%%%%%%%%%%%%%%%%%%%%%%%%%%%%%%%%%%%%%%%%%%%%%%%%%%%%%%%%%%%%%%%%%%%%%%%%%%%%%%%%%%%%%%%%%%%%%%%%%%%%%%%%%%%%%%%%%%%%%%%%%%%%%%%%%%%%%%%%%%%%%%%%%%%%%%%%%%%%%%%%%%%%%%%%%%%%%%%%%%%%%%%%%%%%%%%%%%%%%%%%%%%%%%%%%%
\section{Analysis}\label{sec2}
\begin{figure}[H]
	\centering
	\includegraphics[width=0.8\textwidth]{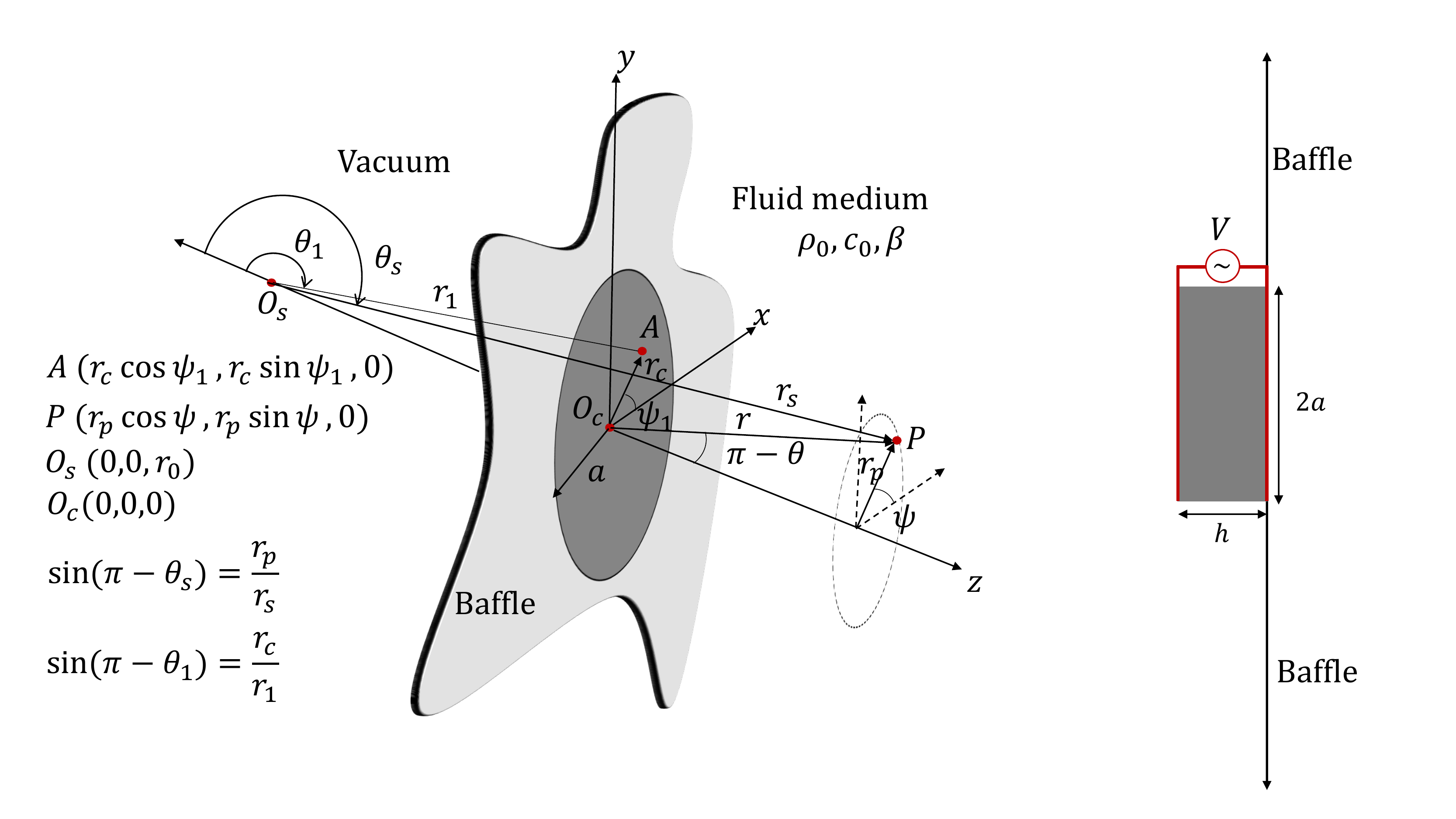}
	\vspace{-15pt}
	\caption{Schematic of the baffled disk with thickness $h$, and radius $a$ in a semi-infinite fluid medium.}\label{fig1}
\end{figure}

An axisymmetric baffled piezoelectric disk with thickness $h$, and radius $a$, in a semi-infinite fluid medium is considered to analyze the nonlinear wave propagation of its generated acoustic wave as shown schematically in Fig. \ref{fig1}. The schematic defines Cartesian $(x,y,z)$, cylindrical $(r_p,\psi,z)$, and spherical coordinate $(r,\theta,\psi)$ systems with the origin at the center of the disk $O_c$. An additional spherical coordinate system $(r_s,\theta_s,\psi)$ with origin, $O_s$, at $z=r_0$ is also defined and used in the analysis. The disk is actuated using a dynamic potential difference $V(t)$ across its flat surfaces at a frequency near that of its thickness mode. The resulting axisymmetric radial and transverse displacements are represented by $\hat{u}(r_p,z,t)$, and $\hat{w}(r_p,z,t)$ respectively. The transverse displacement of the thickness mode is chosen because it has a non-zero mean and is therefore favorable for generating acoustic pressure field \cite{meesala2020acoustic,guo1992finite}.
\noindent The normal velocity continuity condition at the piezo-medium interface yields the necessary radiation boundary condition written as
\begin{equation}
    \frac{\partial }{\partial t}[z-w(r_p,t)] + \nabla\phi(r_p,z,t) \: . \: \nabla [z-w(r_p,t)]=0
    \label{1}
\end{equation}
where $w(r_p,t)=\hat{w}(r_p,z,t)\vert_{z=0}$ and $\phi(r_p,z,t)$ is the velocity potential of the fluid. This boundary condition is an approximation because the displacement, written using the Lagrangian description, would have to be mapped to an equivalent Eulerian description to obtain the exact boundary condition. Still, this approximation is acceptable because $r_p>>>\hat{u}(r_p,z,t)\vert_{z=0}$, except at $r_p=0$ where $\hat{u}(r_p,z,t)\vert_{z=0}=0$, and, consequently, the effects of the approximation are insignificant. Because the modal deformation of the disk can take a complicated shape and does not possess a closed form expression \cite{guo1992finite}, we write $w(r_p,t)$ as
\begin{equation}
w\left(r_p,t\right)=
\begin{cases} 
-\epsilon \left(\frac{c_0}{\omega}\right) \cos(\omega t)G(r_p)e^{i\text{\o}(r_p)} & 0\leq r_p\leq a \\
0 & r_p > a
\end{cases}
\label{2}
\end{equation}
where $\epsilon$, $c_0$, and $\omega$ are respectively the acoustic Mach number, velocity of sound in fluid, and excitation frequency, and $G(r_p)$, and $\text{\o}(r_p)$ are respectively the displacement amplitude and phase as a function of the distance $r_p$.  

The loss-less form of the Wesetervelt equation is used to analyze the nonlinear and non-planar wave propagation of the acoustic pressure $p(r_p,z,t)$. It is written as
\begin{equation}
\nabla^2 p(r_p,z,t)-\frac{1}{c_0^2}\frac{\partial^2 }{\partial t^2}p(r_p,z,t)=-\frac{\beta}{\rho_0c_0^4}\frac{\partial^2 }{\partial t^2}p(r_p,z,t)^2
\label{3}	
\end{equation}
where $\rho_0$ and $\beta$ are respectively the density and the coefficient of nonlinearity of the fluid. It is relevant to note that the loss-less form is valid if the analysis is restricted to distances smaller than $1/\alpha-\overline{x}_{\epsilon}$, where $\alpha$ is the absorption loss parameter of the fluid and $\overline{x}_\epsilon$ is the shock formation distance for a plane wave\cite{muir1980prediction}. To solve equation \ref{3} subjected to the boundary condition given by equation \ref{1}, we asymptotically expand the pressure as
\begin{equation}
p(r_p,z,t,\epsilon)= \epsilon p_1(r_p,z,t) + \epsilon^2 p_2(r_p,z,t) + O(\epsilon^3)
\end{equation} 
and the velocity potential as
\begin{equation}
\phi(r_p,z,t,\epsilon)= \epsilon \phi_1(r_p,z,t) + \epsilon^2 \phi_2(r_p,z,t) + O(\epsilon^3)
\end{equation} 
Substituting the assumed expansions of $p(r_p,z,t)$ and $\phi(r_p,z,t)$ into equations \ref{1}, and \ref{3} yields the $\epsilon-$order equation:
\begin{subequations}
	\begin{equation}
	\frac{1}{c_0^2}\frac{\partial^2}{\partial t^2}p_1(r_p,z,t)-\nabla^2p_1(r_p,z,t)=0, 
	\end{equation}
	and the corresponding boundary condition:
	\begin{equation}
	\left.\frac{\partial }{\partial z}\phi_1(r_p,z,t)\right\vert_{z=0}= c_0\sin(\omega t)G(r_p)e^{i\text{\o}(r_p)},
	\end{equation}
	and the $\epsilon^2-$order equation:
	\begin{equation}
	\frac{1}{c_0^2}\frac{\partial^2}{\partial t^2}p_2(r_p,z,t)-\nabla^2p_2(r_p,z,t)=\frac{\beta}{\rho_0c_0^4}\frac{\partial^2 }{\partial t^2}p_1(r_p,z,t)^2
	\label{6c}
	\end{equation}
\end{subequations}
From Rayleigh's or KH integral, the pressure generated due to a baffled-vibrating surface of surface area $S_0$ and transverse velocity $U_0e^{i\omega t}$ is given by \cite{kinsler1999fundamentals,kim2010sound}
\begin{equation}
p = \frac{1}{2}\frac{i\rho_0\omega}{2\pi}\int\displaylimits_{S_0}\frac{U_0}{R}e^{i\left[\omega t-kR\right]}\,dS_0+c.c
\end{equation}
where $R$ is the distance between the observation point and an infinitesimal point on the surface ($AP$ in Fig. \ref{fig1}) and $k=\omega/c_0$. As such, the $\epsilon-$order solution is given by
\begin{equation}
p_1(r_p,z,t)=\frac{\rho_0\omega c_0}{4\pi}e^{i\omega t}\int\displaylimits_{0}^{2\pi}\int\displaylimits_{0}^{a}\frac{1}{R}e^{i \left[-kR+\text{\o}(r_p)\right]}G(r_p) r_p\,d r_p\,d\psi_1 + c.c
\label{8}
\end{equation}
The $\epsilon-$order solution in equation \ref{8} contains the independent variables $r_p$ and $z$ inside the integral. Hence, a partial differential equation with an integral forcing function needs to be solved to determine the $\epsilon^2-$order solution. Such a solution is not straightforward. To simplify the analysis, we rewrite $p_1(r_p,z,t)$ in terms of an infinite series as suggested by Hasegawa et al. \cite{hasegawa1983new}, using the spherical coordinate system $(r_s,\theta_s,\psi)$ with the origin at $z=r_0$ (see Fig. \ref{fig1}), as 
\begin{subequations}
	\begin{equation}
		p_1(r_s,\theta_s,t)=\frac{\rho_0\omega c_0}{4\pi}e^{i\omega t}\int\displaylimits_{0}^{2\pi}\int\displaylimits_{0}^{a}\frac{1}{R}e^{i \left[-kR+\text{\o}(r_p)\right]}G(r_p) r_p\,d r_p\,d\psi_1 + c.c
		\label{9a}
	\end{equation}
	where 
	\begin{equation}
	    R=\sqrt{r_s^2+r_1^2-2r_sr_1\cos\gamma}
	\end{equation}
	and
	\begin{equation}
	    \cos\gamma=\cos\theta_s\cos\theta_1 + \sin\theta_s\sin\theta_1\cos(\psi-\psi_1).
	\end{equation} 
	Following Hasegawa et al. \cite{hasegawa1983new}, equation \ref{9a} is modified to
	\begin{equation}
	p_1(r_s,\theta_s,t)=-\frac{i\rho_0\omega c_0k}{2}e^{i\omega t}\sum_{n=0}^{\infty}\Delta_n(2n+1) h_n^{(2)}(kr_s)P_n(\cos \theta_s ) + c.c
	\end{equation}
	where 
	\begin{equation}
	    \Delta_n =\int\displaylimits_{r_0}^{r_a}j_n(kr_1)P_n(\cos\theta_1)G(r_1)e^{j\text{\o}(r_1)} r_1\,dr_1,
	\end{equation} $r_1^2=r_p^2 + r_0^2$, and $r_a^2=a^2 + r_0^2$, $P_n$ is the Legendre function and $h_n^{(2)}$ is the spherical Hankel function of the second kind, which is defined by $h_n^{(2)}=j_n-iy_n$ where $j_n$ and $y_n$ are respectively the spherical Bessel functions of the first and second kind. For subsequent analysis, the $\epsilon-$order solution is converted from exponential to trigonometric form and written as 
	\begin{multline}
	p_1(r_s,\theta_s,t)=
	\rho_0kc_0\omega \sum_{n=0}^{n=\infty}(2n+1)P_n(\cos\theta_s)\left(\Delta^{(c)}_n\left[\sin (\omega t )j_n(kr_s) - \cos (\omega t )y_n(kr_s)\right]\right.\\\left.
	+\Delta^{(s)}_n\left[\cos (\omega t )j_n(kr_s) + \sin (\omega t )y_n(kr_s)\right]\right)
	\label{9f}
	\end{multline}
	where
	\begin{equation}
	   \Delta^{(c)}_n =\int\displaylimits_{r_0}^{r_a}j_n(kr_1)P_n(\cos\theta_1)G\cos(\text{\o}(r_1)) r_1\,dr_1
	\end{equation}
	\begin{equation}
	  \Delta^{(s)}_n =\int\displaylimits_{r_0}^{r_a}j_n(kr_1)P_n(\cos\theta_1)G\sin(\text{\o}(r_1)) r_1\,dr_1
	\end{equation}
\end{subequations}
Substituting the $\epsilon-$order solution in the $\epsilon^2-$order equation (\ref{6c}) yields
\begin{subequations}
    \begin{multline}
\frac{1}{c_0^2}\frac{\partial^2}{\partial t^2}p_2(r_s,\theta_s,t)-\nabla^2p_2(r_s,\theta_s,t)=
-2\sum_{n=0}^{\infty}\sum_{m=0}^{\infty}(2n+1)(2m+1)k^4\beta\rho_0\omega^2P_n(\cos\theta_s)P_m(\cos\theta_s)\\
\times\left(\frac{ }{ }E_1\cos\left(2\omega t\right)+E_2\sin\left(2\omega t\right)\frac{ }{ }\right)
\label{10a}
\end{multline}
where
\begin{multline}
   E_1 = \left(\Delta_n^{(c)}\Delta_m^{(c)}-\Delta_n^{(s)}\Delta_m^{(s)}\right)\left(\frac{ }{ }j_n(kr_s)j_m(kr_s)-y_n(kr_s)y_m(kr_s)\frac{ }{ }\right)\\
   +\left(\Delta_n^{(c)}\Delta_m^{(s)}+\Delta_n^{(s)}\Delta_m^{(c)}\right)\left(\frac{ }{ }j_m(kr_s)y_n(kr_s)+j_n(kr_s)y_m(kr_s)\frac{ }{ }\right), \text{  and}
\end{multline}
\begin{multline}
E_2=\left(\Delta_n^{(c)}\Delta_m^{(c)}-\Delta_n^{(s)}\Delta_m^{(s)}\right)\left(\frac{ }{ }j_m(kr_s)y_n(kr_s)+j_n(kr_s)y_m(kr_s)\frac{ }{ }\right)\\
-\left(\Delta_n^{(c)}\Delta_m^{(s)}+\Delta_n^{(s)}\Delta_m^{(c)}\right)\left(\frac{ }{ }j_n(kr_s)j_m(kr_s)-y_n(kr_s)y_m(kr_s)\frac{ }{ }\right)
\end{multline}
\end{subequations}
To obtain the analytical expression for the $\epsilon^2-$order solution, the product of Legendre functions is rewritten as\cite{kelly1980non}
\begin{equation}
P_n(\cos \theta_s) P_m(\cos \theta_s)=\sum_{q=0}^{p+n}\kappa_{qnm}P_q(\cos\theta_s)
\label{11}
\end{equation}
where the coefficient $\kappa_{qnm}$ is determined from the orthogonality condition of the Legendre-functions as
\begin{equation}
	\kappa_{qnm}=\frac{2q+1}{2}\int\displaylimits_{-1}^{1}P_q(x)P_n(x) P_m(x)\,dx
\end{equation}
Substituting equation \ref{11} into equation \ref{10a}, we obtain
\begin{multline}
\frac{1}{c_0^2}\frac{\partial^2}{\partial t^2}p_2(r_s,\theta_s,t)-\nabla^2p_2(r_s,\theta_s,t)=
-2\sum_{n=0}^{\infty}\sum_{m=0}^{\infty}\sum_{q=0}^{n+m}(2n+1)(2m+1)k^4\beta\rho_0\omega^2\kappa_{qnm}P_q(\cos\theta_s)\\
\times\left(\frac{ }{ }E_1\cos\left(2\omega t\right)+E_2\sin\left(2\omega t\right)\frac{ }{ }\right)
\label{13}
\end{multline}
As detailed in Appendix \ref{appa}, the solution of $\epsilon^2-$order equation (\ref{13}), $p_2(r_s,\theta_s,t)$, is obtained using the separation of variables. The final solution of the pressure is then written as
\begin{multline}
p(r_s,\theta_s,t,\epsilon)=
\epsilon\rho_0kc_0\omega \sum_{n=0}^{n=\infty}\Delta^{(c)}_n(2n+1)P_n(\cos\theta_s)\left[\sin (\omega t )j_n(kr_s) - \cos (\omega t )y_n(kr_s)\right]\\
+\epsilon\rho_0kc_0\omega \sum_{n=0}^{n=\infty}\Delta^{(s)}_n(2n+1)P_n(\cos\theta_s)\left[\cos (\omega t )j_n(kr_s) + \sin (\omega t )y_n(kr_s)\right]
+\epsilon^2p_2(r_s,\theta_s,t)+O(\epsilon^3)
\label{14}
\end{multline}
The $\epsilon-$order and $\epsilon^2-$order terms in equation \ref{14} decay at different rates with respect to $r_s$. If left untreated, the $\epsilon^2-$order terms can become significant and contradict the scaling condition that $\epsilon-$order terms $>>$ $\epsilon^2-$order terms. To solve this contradiction and eliminate terms leading to this contradiction, we implement the method of renormalization and introduce the coordinate transformation \cite{kelly1980non}  
\begin{equation}
r_s=\eta+\epsilon f(\eta,\theta_s,t)
\end{equation}
Substituting the above transformation into equation \ref{14} and applying the Taylor expansion yields
\begin{equation}
p(r_s,\theta_s,t,\epsilon)=\epsilon p_1(\eta,\theta_s,t)
+\epsilon^2f(\eta,\theta_s,t)\left.\frac{\partial }{\partial r_s}p_1(r_s,\theta_s,t)\right\vert_{r_s=\eta}+\epsilon^2p_2(\eta,\theta_s,t)+O(\epsilon^3)
\label{16}
\end{equation}
From equation \ref{16}, the necessary condition to eliminate the secular terms is
\begin{equation}
f(\eta,\theta_s,t)\left.\frac{\partial}{\partial r_s}p_1(r_s,\theta_s,t)\right\vert_{r_s=\eta}+p_2(\eta,\theta_s,t)=0
\label{17}
\end{equation}
The transformation $f$ becomes singular unless there is a particular relation between the phases of $p_2$ and $\left.\partial p_1/\partial r_s\right.\vert_{r_s=\eta}$. To examine the phase relation, we represent $p_2$ and $\left.\partial p_1/\partial r_s\right.\vert_{r_s=\eta}$ respectively as $p_2(\eta,\theta_s,t)=\overline{p}_2(\eta,\theta_s)\cos(2\omega t -2 k \eta + \text{\o}_{p{_2}})$ and $\left.\partial p_1/\partial r_s\right.\vert_{r=\eta}=\overline{p}'_1(\eta,\theta_s)\cos(\omega t - k \eta + \text{\o}_{p'{_1}})$ where $\overline{p}_2(\eta,\theta_s)$, $\text{\o}_{p_{2}}(\eta,\theta_s)$, $\overline{p}'_1(\eta,\theta_s)$, and $\text{\o}_{p'{_1}}(\eta,\theta_s)$ are respectively the amplitude and phase of $p_2$ and amplitude and phase of $\left.\partial p_1/\partial r_s\right.\vert_{r_s=\eta}$. From equation \ref{17}, $f(\eta,\theta_s,t)$ is then determined as
\begin{multline}
f(\eta,\theta_s,t) =
\frac{2\overline{p}_2(\eta,\theta_s)}{\overline{p}'_1(\eta,\theta_s)}\cos(\pi/4- \omega t +  k \eta - \text{\o}_{p{_2}}(\eta,\theta_s)/2)\left[\frac{ }{ }\cos(- \pi/4 + \text{\o}_{p'{_1}}- \text{\o}_{p{_2}}(\eta,\theta_s)/2) \right.\\\left.
+ \tan(\pi/2 - \omega t +  k \eta -\text{\o}_{p'{_1}})\sin(- \pi/4 + \text{\o}_{p'{_1}}(\eta,\theta_s)- \text{\o}_{p{_2}}(\eta,\theta_s)/2)\frac{ }{ }\right]
\label{18}
\end{multline}
It is evident from equation \ref{18} that singularities arise in $f$ because of the tangent functions, unless $\text{\o}_{p'{_1}}- \text{\o}_{p{_2}}/2=\pi/4$. However, $\text{\o}_{p'{_1}}$ and $\text{\o}_{p{_2}}$ do not always hold such a relation. To eliminate this singularity, we rewrite $p_2(\eta,\theta_s,t)$ as
\begin{multline}
p_2(\eta,\theta_s,t)=2	\frac{\overline{p}_2(\eta,\theta_s)}{\overline{p}'_1(\eta,\theta_s)}\left.\frac{\partial p_1}{\partial r}\right\vert_{r=\eta}\cos\left(\frac{ }{ }\omega t - k\eta + \text{\o}_{p{_2}}(\eta,\theta_s)-\text{\o}_{p'{_1}}(\eta,\theta_s)\frac{ }{ }\right)\\
-	\overline{p}_2(\eta,\theta_s)\cos\left(\frac{ }{ }\text{\o}_{p{_2}}(\eta,\theta_s)-2\text{\o}_{p'{_1}}(\eta,\theta_s)\frac{ }{ }\right)
\label{19}
\end{multline} 
Then, eliminating the first part on the right hand side of equation \ref{19} and using the remaining time-independent term as a feedback error to the $\epsilon-$order solution yields
\begin{equation}
	p(r_s,\theta_s,t,\epsilon)=\epsilon p_1(\eta,\theta_s,t)-\epsilon^2\overline{p}_2(\eta,\theta_s)\cos(\text{\o}_{p{_2}}(\eta,\theta_s)-2\text{\o}_{p'{_1}}(\eta,\theta_s))
	\label{20}
\end{equation}
where 
\begin{equation}
	r_s=\eta+\epsilon f(\eta,\theta_s,t)
	\label{21}
\end{equation}
\begin{equation}
    f(\eta,\theta_s,t) = -2\frac{\overline{p}_2(\eta,\theta_s)}{\overline{p}'_1(\eta,\theta_s)}\cos(\omega t - k\eta + \text{\o}_{p{_2}}(\eta,\theta_s)-\text{\o}_{p'{_1}}(\eta,\theta_s))
    \label{22}
\end{equation}
From equation \ref{22}, it is noted that the transformation $f$ is independent of $\epsilon$. This is a powerful consequence of the method of renormalization as once $f$ is determined from $p_1$ and $p_2$, which are again independent of $\epsilon$, it can be used for any value of $\epsilon$. 

Equations \ref{20} - \ref{22} constitute the solution of the nonlinear wave propagation of the acoustic pressure generated by a vibrating disk with transverse excitation according to equation \ref{2}. For a given $r_s$, $\theta_s$, and $t$, one needs to first evaluate $f$ from equation \ref{22} and then solve for $\eta$ using equation \ref{21}. The value of $\eta$ can then be used to determine the nonlinear pressure from equation \ref{20}. As the wave propagates in the medium, equation \ref{21} will eventually yield multiple solutions for $\eta$ due to the cumulative nature of the nonlinearity \cite{coulouvrat1991analytical,hamilton1998nonlinear}. The first location where multiple solutions occur or when the $\partial p/\partial r_s=\infty$ is the shock location \cite{nayfeh1978non,ginsberg1978propagation,ginsberg1978propagation2,kelly1980non,hamilton1998nonlinear}. Beyond this location, the solution is not valid and a shock fitting criteria such as equal-area rule should be used \cite{hamilton1998nonlinear,nayfeh2008nonlinear}.
%%%%%%%%%%%%%%%%%%%%%%%%%%%%%%%%%%%%%%%%%%%%%%%%%%%%%%%%%%%%%%%%%%%%%%%%%%%%%%%%%%%%%%%%%%%%%%%%%%%%%%%%%%%%%%%%%%%%%%%%%%%%%%%%%%%%%%%%%%%%%%%%%%%%%%%%%%%%%%%%%%%%%%%%%%%%%%%%%%%%%%%%%%%%%%%%%%%%%%%%%%%%%%%%%%%%%%%%%%%%%%%%%%%%%%%%%%%%%%%%%%%%%%%%%%%%%%%%%%%%%%%%%%%%%%%%%%%%%%%%%%%%%%%%%%%%%%%%%%%%%%%%%%%%%%%%%%%%%%%%%%%%%%%%%%%%%%%%
\section{Validation}
The efficacy of the analysis presented in the previous section to predict the nonlinear wave propagation and shock formation is assessed by comparing its predictions with those from (a) higher fidelity Finite Element (FE) simulations and (b) previously published experimental results.

\subsection{Comparison with Finite Element simulations}
A piezoelectric disk with radius $a=5$ mm and thickness $h=2$ mm made of PZT$-5$H material and submerged in water was considered for validation using numerical simulations. The material and acoustic properties of the disk and fluid are presented in Table \ref{tab1}. The corresponding Rayleigh far-field distance is defined by $r_{far}=1/2ka^2=53.3$ mm. We performed a series of axisymmetric frequency and time-domain FE simulations in COMSOL Multiphysics to obtain the acoustic radiation characteristics of the baffled-disk configuration. A quarter-circular fluid domain of radius $r_{f}$ was considered in the simulations. A spherical wave radiation boundary condition at the outer boundary was used to simulate an infinitely propagating wave.  
\begin{table}[H]
	\begin{center}       
		\caption{Material properties of the disk and fluid. $C^E$, $e$, and $\varepsilon^S$ are respectively the the stiffness at constant electric field, coupling
        parameter, and relative permitivity at constant strain matrices.} 
		\begin{tabular}{|l|l|}
			\hline
			\rule[-1ex]{0pt}{3.5ex} \textbf{Parameter [units]} & \textbf{Value}  \\
			\hline
			\rule[-1ex]{0pt}{3.5ex} Density of the disk, $\rho$ [kg/m$^{3}$] & $7500$  \\
			\hline
			\rule[-1ex]{0pt}{3.5ex} $C^E_{11}$ [GPa] &   $127.21$\\
			\hline
			\rule[-1ex]{0pt}{3.5ex} $C^E_{12}$ [GPa] &   $80.21$\\
			\hline
			\rule[-1ex]{0pt}{3.5ex} $C^E_{13}$ [GPa] &   $84.67$\\
			\hline
			\rule[-1ex]{0pt}{3.5ex} $C^E_{33}$ [GPa] &   $117.44$\\
			\hline
			\rule[-1ex]{0pt}{3.5ex} $C^E_{44}$ [GPa] &   $22.99$\\
			\hline
			\rule[-1ex]{0pt}{3.5ex} $e_{31}$ [C/m$^2$] &   $-6.63$\\
			\hline
			\rule[-1ex]{0pt}{3.5ex} $e_{33}$ [C/m$^2$] &   $23.24$\\
			\hline
			\rule[-1ex]{0pt}{3.5ex} $e_{24}$ [C/m$^2$] &   $17.03$\\
			\hline
			\rule[-1ex]{0pt}{3.5ex} $\varepsilon^S_{11}$ &   $1704.4$\\
			\hline
			\rule[-1ex]{0pt}{3.5ex} $\varepsilon^S_{33}$ &   $1433.6$\\
			\hline
			\rule[-1ex]{0pt}{3.5ex} Density of the fluid, $\rho_0$ [kg/m$^3$]&   $999.6$\\
			\hline
			\rule[-1ex]{0pt}{3.5ex} Velocity of sound in the fluid, $c_0$ [m/s]&   $1481.44$\\
			\hline
			\rule[-1ex]{0pt}{3.5ex} Nonlinear parameter of the fluid, $\beta$ &   $3.6$\\
			\hline
		\end{tabular}
		\label{tab1}
	\end{center}
\end{table}
\subsubsection{Linear acoustic radiation}
In both analysis and simulations, the material and geometric nonlinearities of the disk are neglected. Furthermore, because of the presence of a baffle, only the transverse displacement of the top surface of the disk is required to determine its acoustic radiation. In the simulations, we extracted this displacement from a linear FE simulation and use it as a boundary condition in the nonlinear FE simulation for computational efficiency, realized by eliminating the multiphysics coupling between the piezoelectric and fluid domains.

\begin{figure}[!]
	\centering
	\includegraphics[width=1\textwidth]{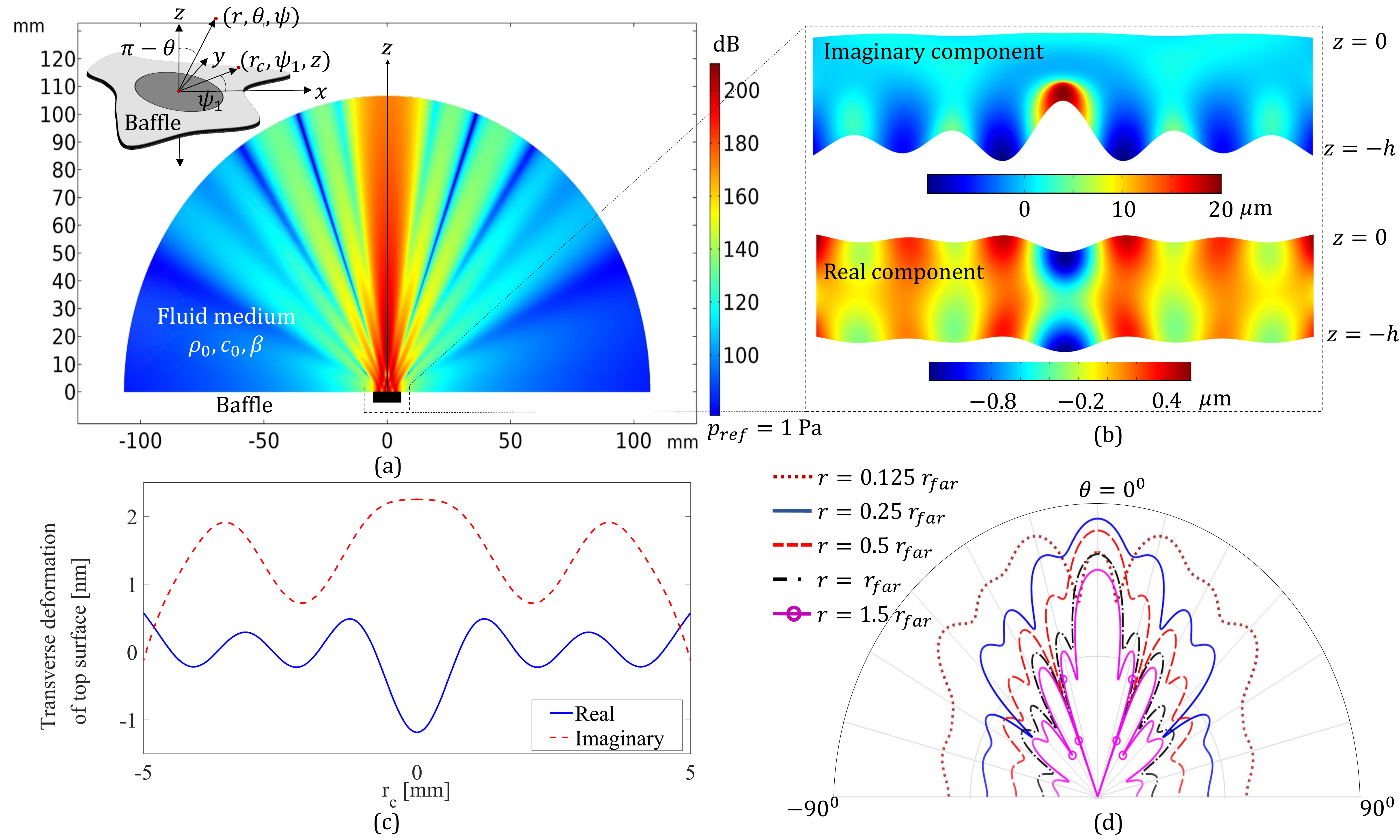}
	\vspace{-20pt}
	\caption{(a) The sound pressure level ($p_{ref}=1$ Pa) in the fluid domain, (b) real and imaginary components of the deformation of the disk, (c) transverse displacement of the top surface of the disk, and (d) beam (directivity) pattern at different radial distances from the origin, for $\epsilon=10^{-5}$ at an excitation frequency of $1.005$ MHz.}\label{fig2}
\end{figure}

Considering a maximum mesh size of $\lambda_{f_0}/6$ in the fluid domain and $20$ elements each along the radius and thickness of the disk, we solved the linear problem using frequency domain FE solver and determined that the thickness mode resonant frequency of the disk is $f_0=1.005$ MHz, where $\lambda_{f_0}=c_0/f_0$. The response and radiation characteristics of the disk for an excitation voltage amplitude corresponding to $\epsilon=10^{-5}$ and an excitation frequency of $f_0$ are presented in Fig. \ref{fig2}. In particular, Figs. \ref{fig2}a-\ref{fig2}d show respectively the sound pressure level ($p_{ref}=1$ Pa) in the fluid domain, deformation of the disk, transverse displacement of the top surface of the disk, and beam pattern at different radial distances from the origin (see schematic in Fig. \ref{fig2}a). The abnormal higher deformation amplitude on the bottom surface compared to the top surface of the disk in Fig. \ref{fig2}b is due to the absence of radiation damping, unlike the top surface. From Fig. \ref{fig2}d, we note that except for $r=0.125r_{far}$, the pressure along the $z-$axis is maximum for any $r$ value. Consequently, we expect the shock to occur first on the $z-axis$. We also note that $r=0.125r_{far}$ is a local minimum, similar to those observed in the near-field of axial pressure generated by a piston \cite{nachef1995investigation,kinsler1999fundamentals,khokhlova2001numerical}. Because we expect the shock to occur along the $z-axis$ first, we limit further analysis only to the axial pressure distribution. 

\subsubsection{Nonlinear axial acoustic wave propagation}\label{sec3p1p2}
\begin{figure}[!]
	\begin{center}
		\subfigure[\label{fig3a}] {\includegraphics[width=0.49\textwidth]{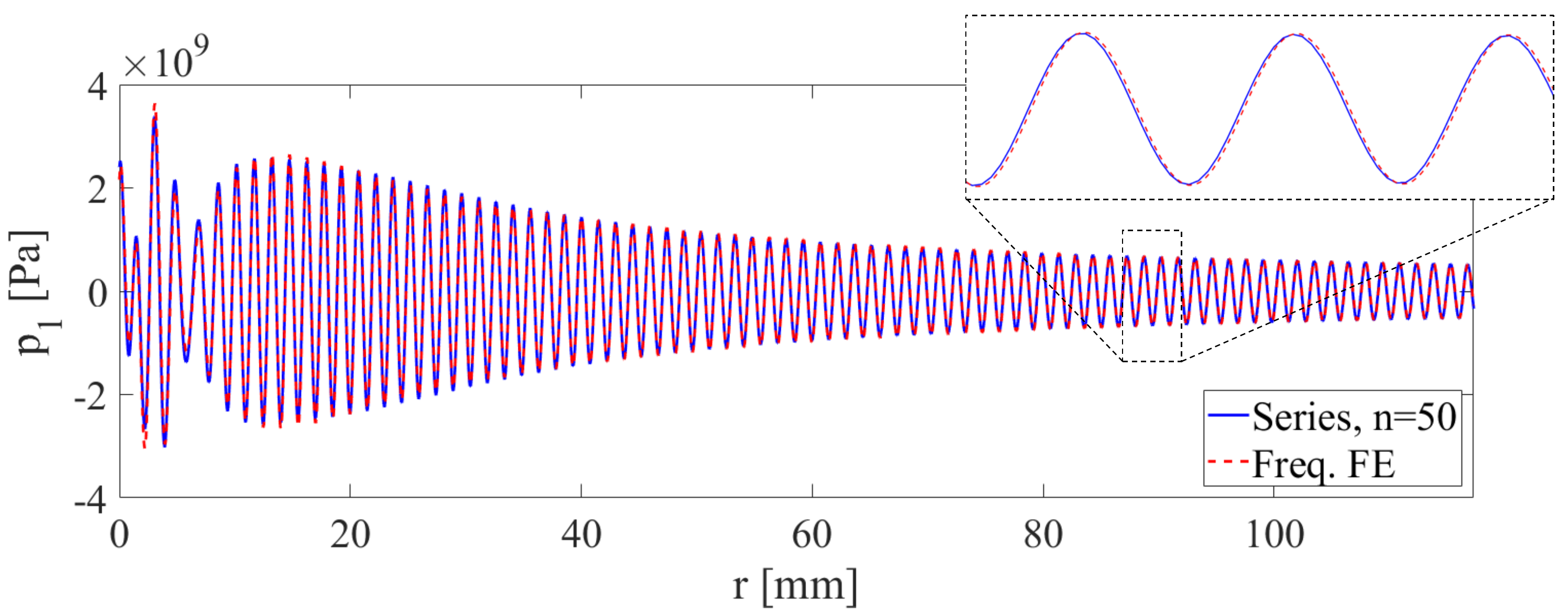}}
		\hspace{0.001in}
		\subfigure[\label{fig3b}] {\includegraphics[width=0.49\textwidth]{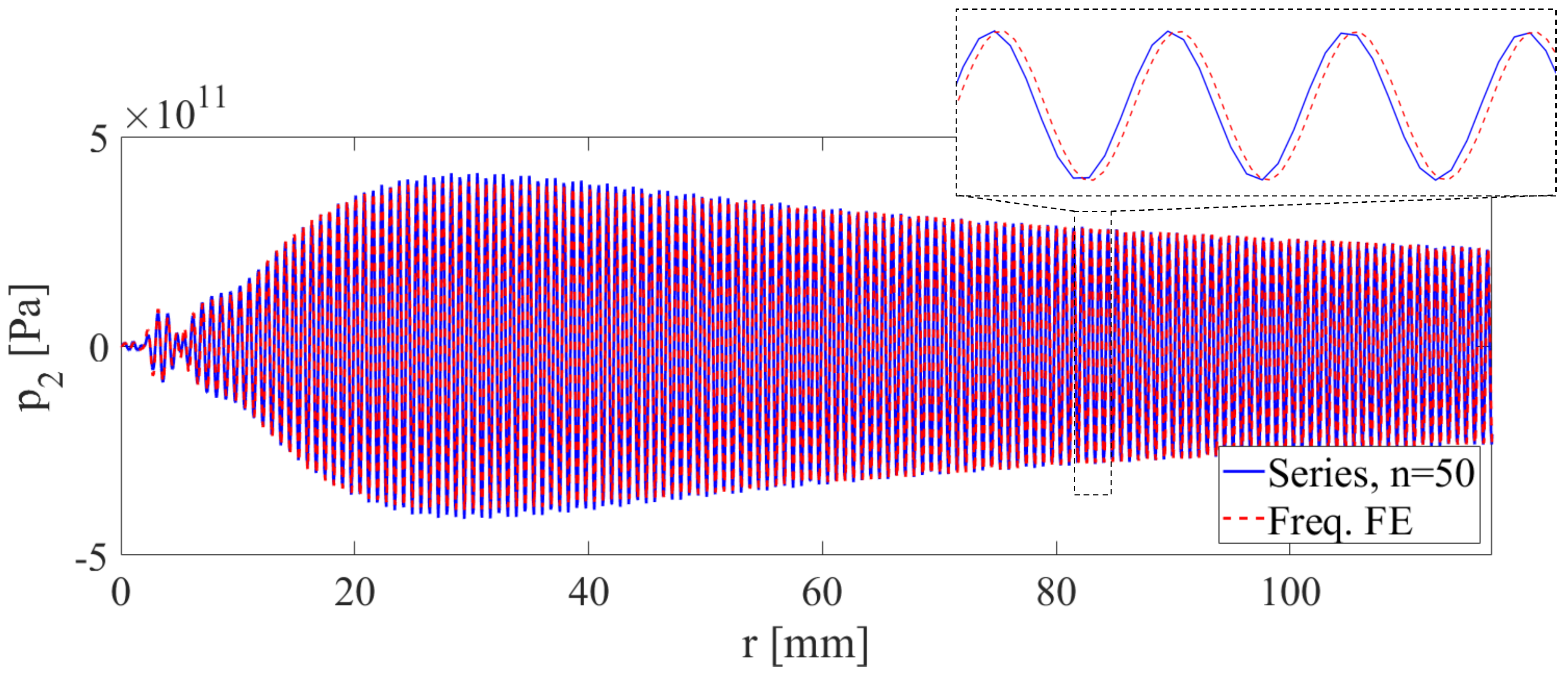}}
	\end{center}
	\vspace{-20pt}
	\caption{Comparison of axial (a) $\epsilon-$order solution $p_1$, and (b) $\epsilon^2-$order solution $p_2$, determined by using the analysis developed for $n=50,r_0=-50/k$ and by a frequency domain FE solver. The analytical solution is labeled as Series, $n=50$ and FE solution is labeled as Freq. FE. On $z-$axis, $r_s=r-r_0$ and $\theta=\theta_s=\pi$. }
\end{figure}

Next, the transverse displacement presented in Fig. \ref{fig2}c is used to obtain axial $p_1$ and $p_2$ according to the solution in section \ref{sec2}. Alternatively, $p_1$ and $p_2$ could be determined by solving the $\epsilon-$order and $\epsilon^2-$order equations and $\epsilon-$order boundary condition in a frequency domain FE solver. A pressure release boundary condition on the disk should be used to solve the $\epsilon^2-$order equations to ensure that $p_2$ vanishes on the surface of the disk. We note that $p_2$ evaluated using the two approaches should differ only by the non-secular terms (NST) that were neglected in the analysis. The axial $p_1$ and $p_2$ evaluated from the model for $n=50, r_0=-50/k$, labeled as Series $n=50$, and by the frequency domain FE simulations are presented respectively in Fig. \ref{fig3a} and Fig. \ref{fig3b}. A mesh size of $\lambda_{f_0}/12$ was used in the FE simulations. Fig. \ref{fig3a} shows an excellent agreement between $p_1$ evaluated by the analysis and numerical simulations except at close distances to the disk. The disagreement in this region is due to limiting the series and can be minimized by choosing higher values of $n$ and $r_0$. From Fig. \ref{fig3b}, we note a close agreement in the amplitude and a small phase shift. We attribute this shift to errors induced by choosing a finite value of $n$, and to the contribution of the NST. It is relevant to point out that $p_1$ and $p_2$ determined using FE are more accurate and that eliminating $p_2$ that includes NST in the method of the renormalization will result in a transformation that will account for the complete particular solution of the $\epsilon^2-$order equation. As for computational cost, the evaluation of $p_1$ and $p_2$ from the analysis requires the analytical expressions of integrals presented in Appendix \ref{appa} and storing, reusing, and manipulation of these expressions, which can become computationally expensive, especially for higher values of $n$. On the other hand, the linear frequency domain FE simulations to determine $p_1$ and $p_2$ is more efficient as it involves a matrix inversion. Moreover, the FE solver can be easily extended to complicated geometries.

Having determined $p_1$ and $p_2$, we compare next the nonlinear wave propagation predicted using equations \ref{20} - \ref{22} with that determined from a nonlinear time-domain FE simulation for three different excitation levels, namely $\epsilon=5\times10^{-3}$, $\epsilon=2.5\times10^{-3}$, and $\epsilon=1.25\times10^{-3}$. Towards that objective, we constructed a fluid domain of $r_f=2.25\times\overline{x}_{\epsilon}$ that is excited by the transverse displacement presented in Fig. \ref{fig2}c. A maximum mesh size of $\lambda_{f_0}/36$ was selected so that the response at $6f_0$ can be captured. For the sake of completeness, Westervelt equation with dissipation was simulated in time-domain using COMSOL Multiphysics by choosing the diffusivity of the sound as $\delta=1.34\times10^{-6}$ m$^2$/s. However, the absorption is not 
expected to have a significant impact on the response because the characteristic absorption distance ($1/\alpha=2c_0^3/(\delta\omega^2)$) for $f_0$ is about $120$ m, which is significantly larger than 
$2.25\times \overline{x}_{1.25\times10^{-3}}(\approx0.1$ m).
\begin{figure}[!]
	\begin{center}
		\subfigure[\label{fig4a}] {\includegraphics[width=0.8\textwidth]{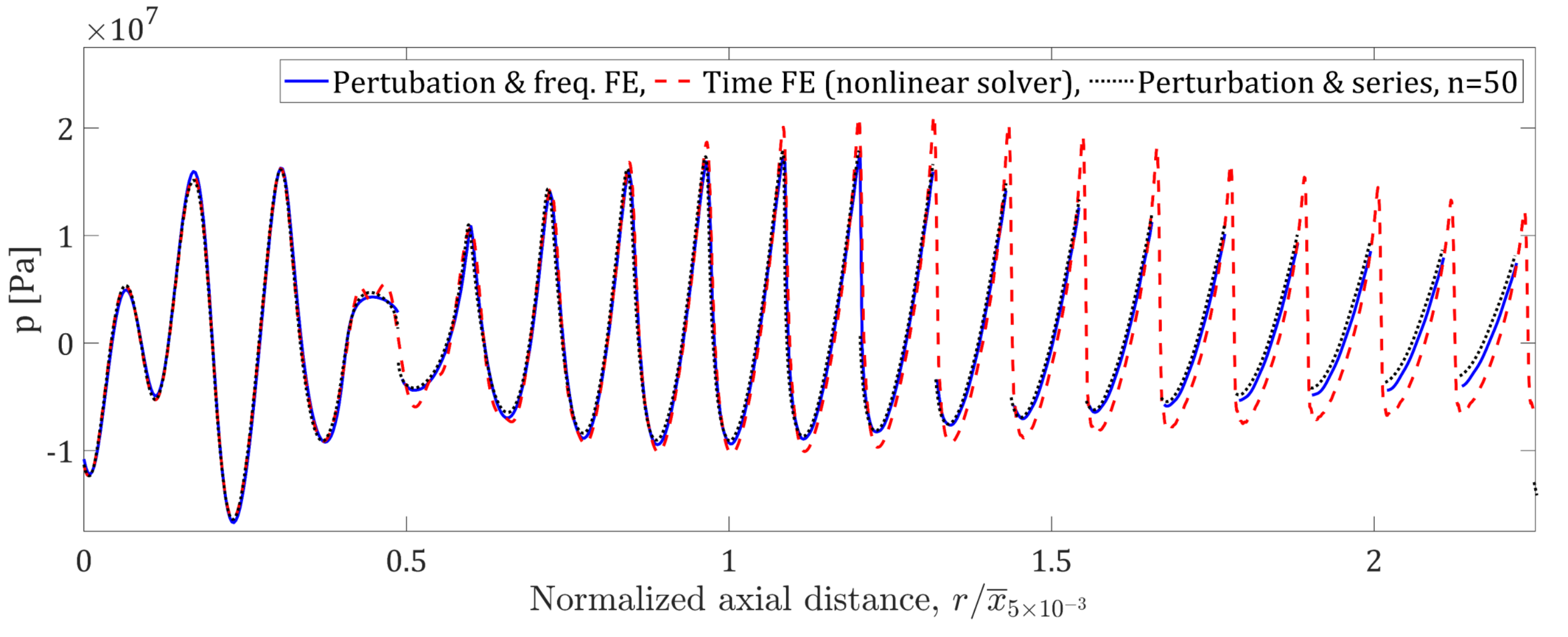}}
		\subfigure[\label{fig4b}] {\includegraphics[width=0.328\textwidth]{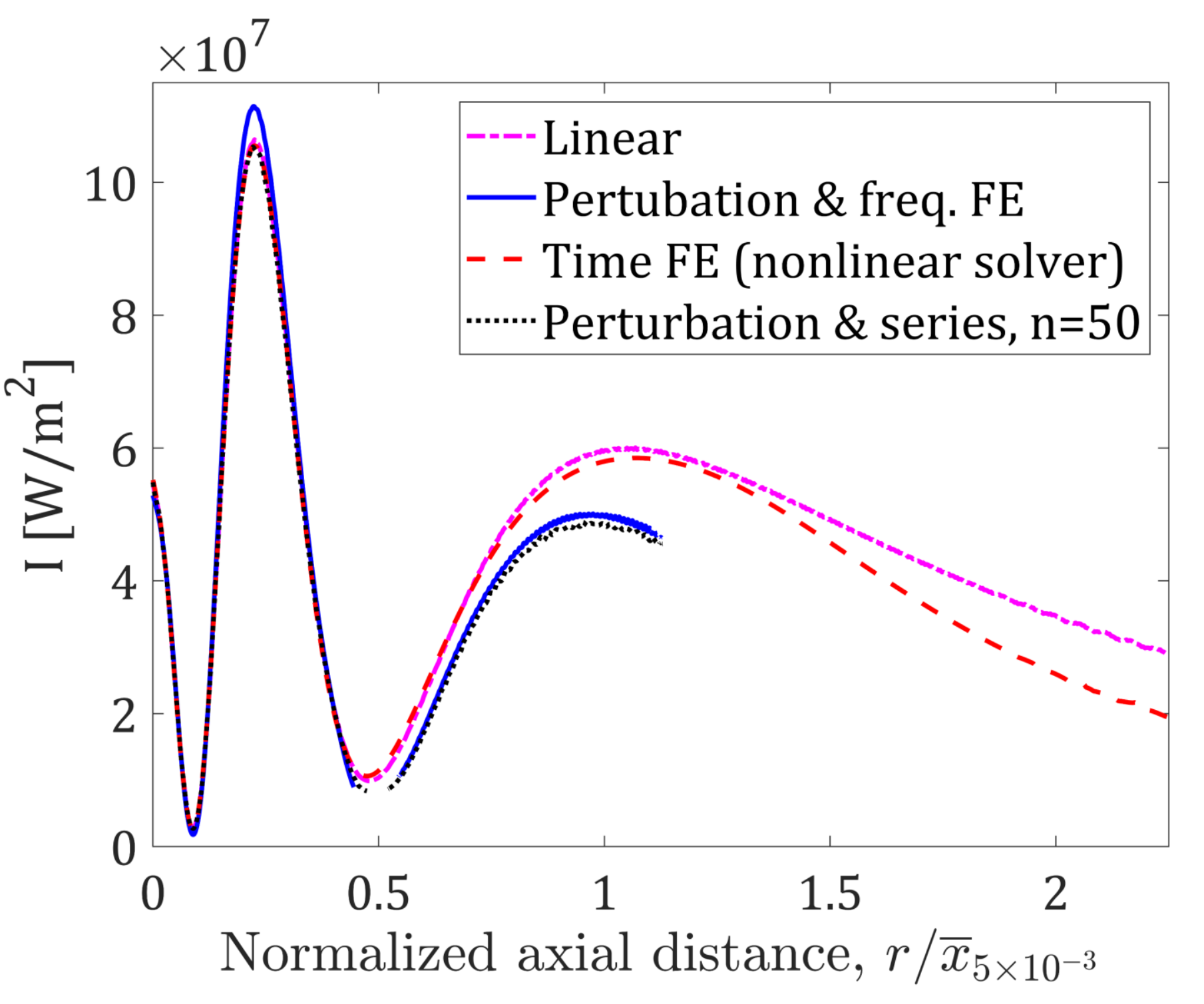}}
		\subfigure[\label{fig4c}] {\includegraphics[width=0.328\textwidth]{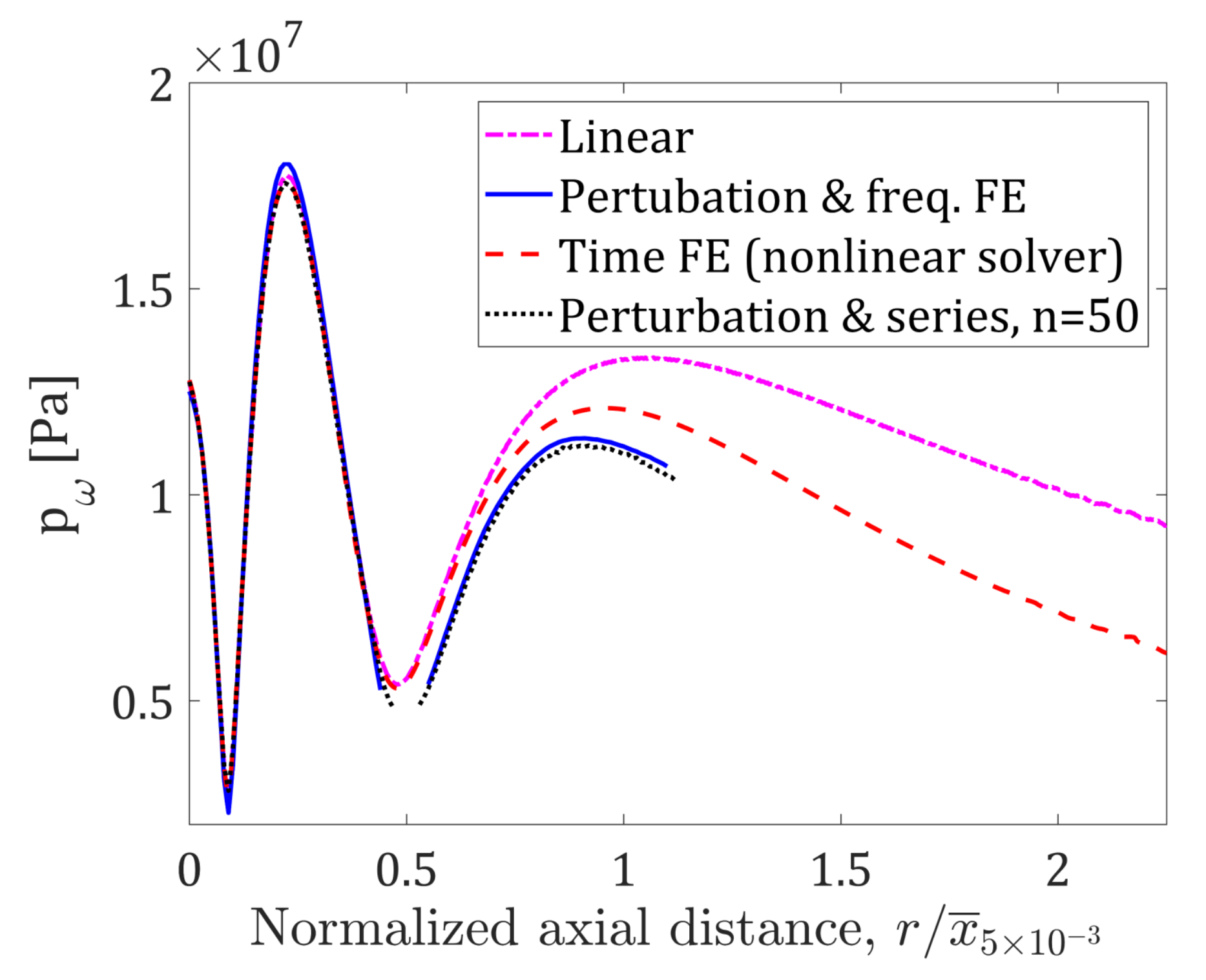}}
		\subfigure[\label{fig4d}] {\includegraphics[width=0.328\textwidth]{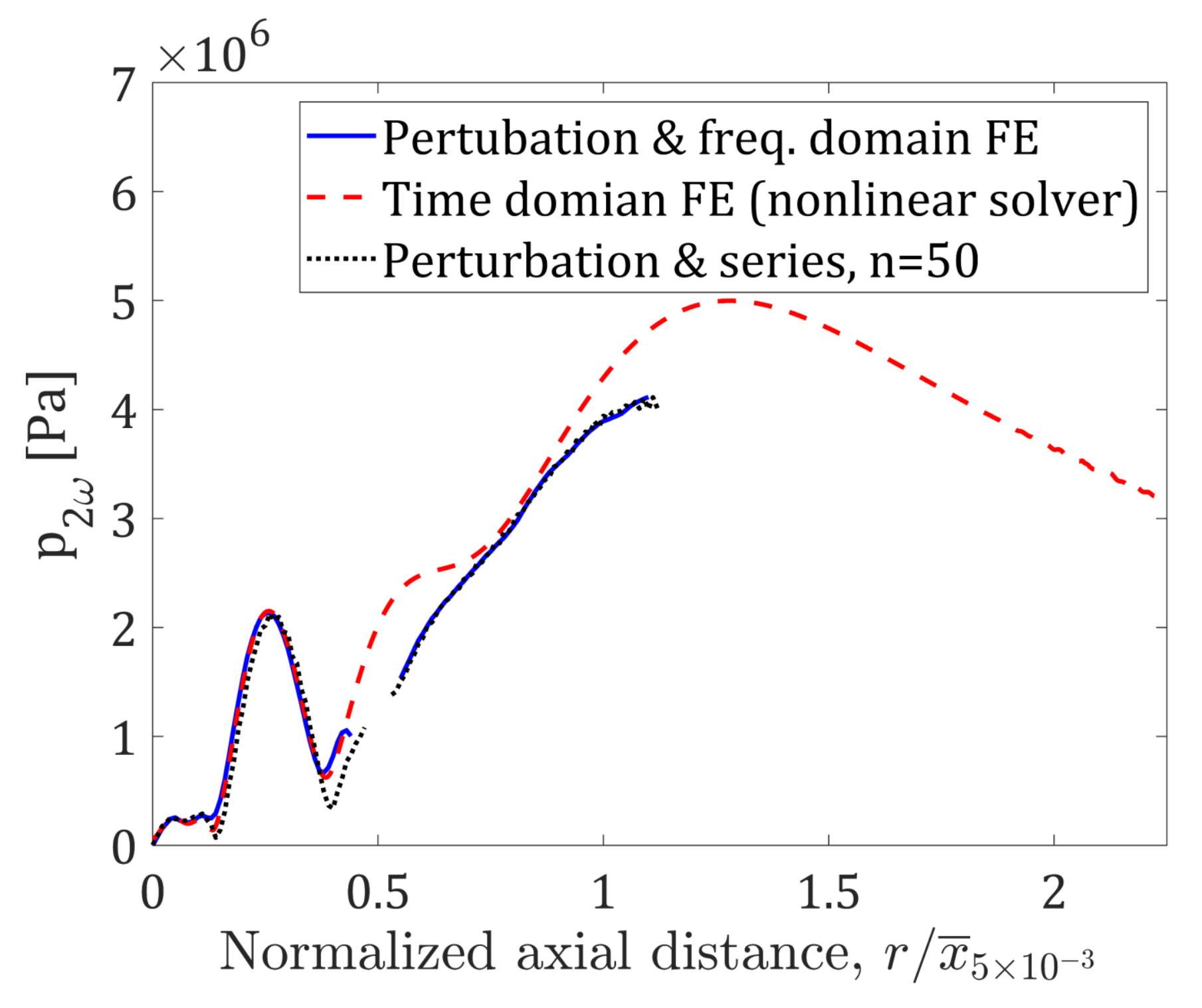}}
	\end{center}
	\vspace{-20pt}
	\caption{Comparison of steady-state axial (a) pressure waveform at time $t=1/(2f_0)$, (b) mean intensity, (c) component of the pressure at $\omega=2\pi f_0$, and (d) component of the pressure at $2\omega$ obtained from time-domain FE simulation and from the model developed when $p_1$ and $p_2$ are obtained from the analysis (Perturbation \& series, $n=50$) and from frequency domain simulations (Perturbation \& freq. FE) for $\epsilon=5\times10^{-3}$ and an excitation frequency $f_0=1.005$ MHz. The axial distance is normalized with the corresponding characteristic shock formation distance $\overline{x}_{5\times 10^{-3}}=13.03$ mm.}
\end{figure}

\begin{figure}[!]
	\begin{center}
		\subfigure[\label{fig5a}] {\includegraphics[width=0.4\textwidth]{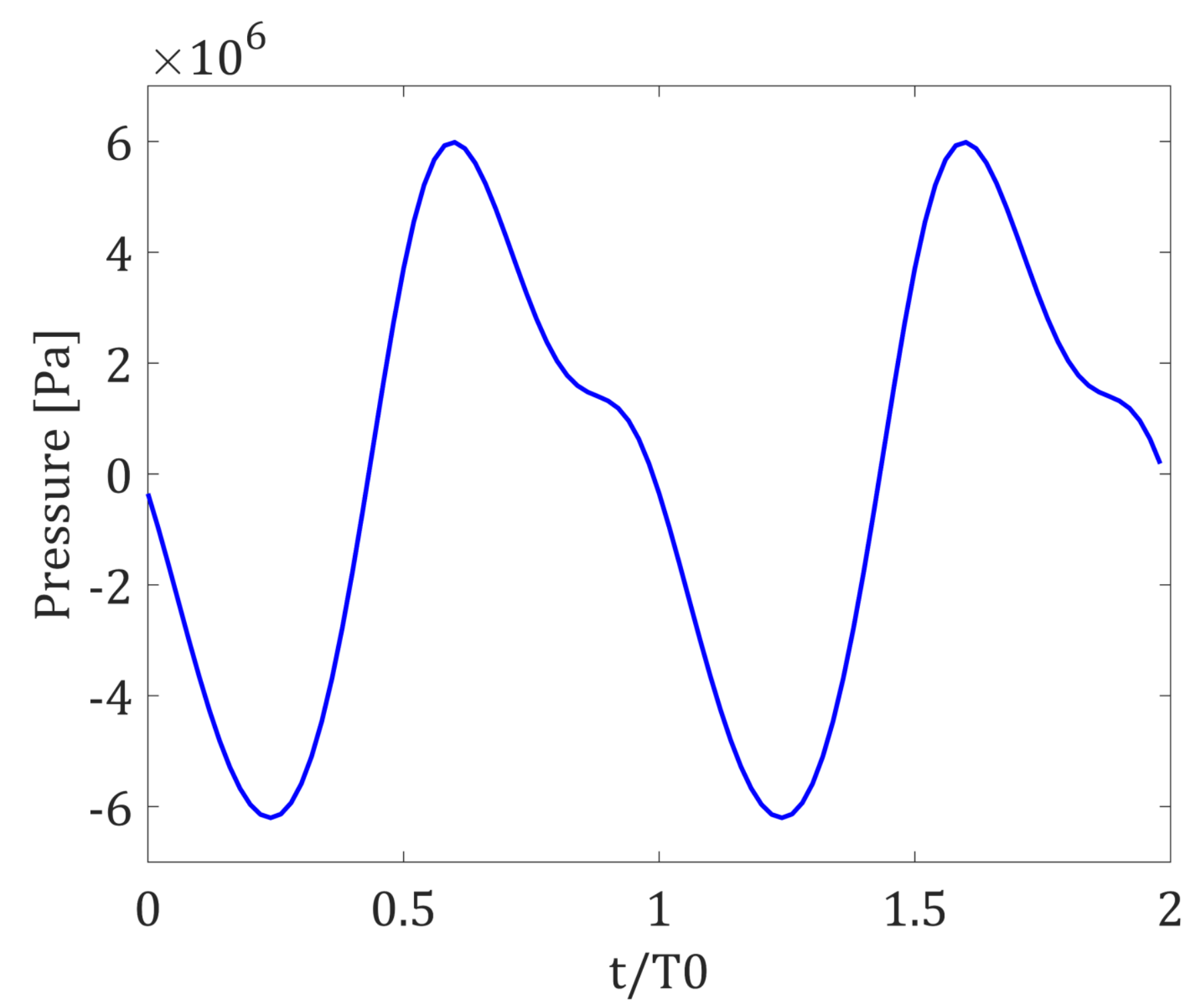}}
		\hspace{0.2in}
		\subfigure[\label{fig5b}] {\includegraphics[width=0.4\textwidth]{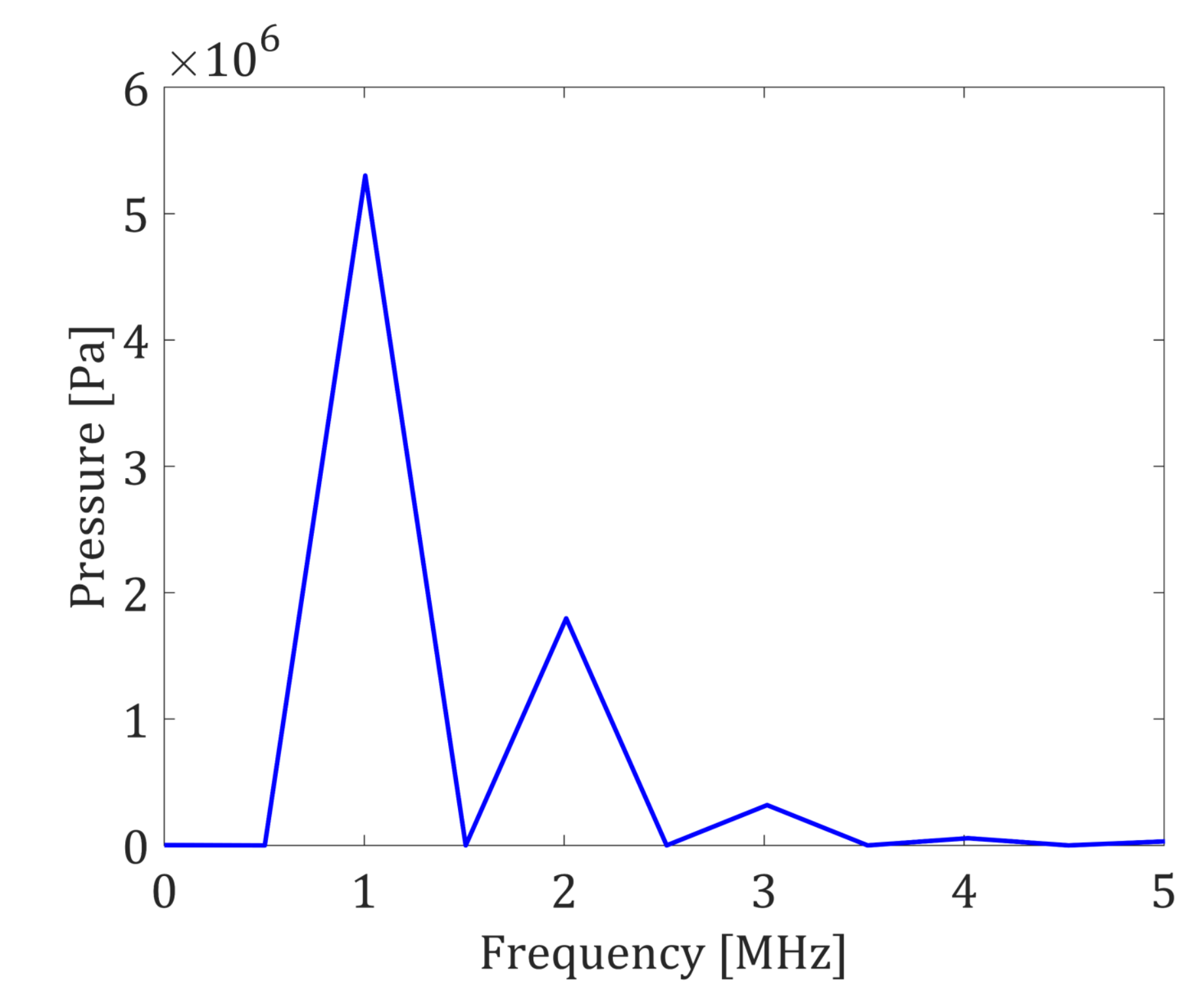}}
	\end{center}
	\vspace{-20pt}
	\caption{The (a) time series and (b) amplitude Fourier coefficients of the pressure at $r/\overline{x}_{5\times10^{-3}}=0.48$ on $z-$axis for $\epsilon=5\times10^{-3}$ and excitation frequency $f_0=1.005$ MHz obtained from a time-domain FE simulation. Here, $\overline{x}_{5\times 10^{-3}}=13.03$ mm.  }
\end{figure}

The steady-state response characteristics for $\epsilon=5\times10^{-3}$ obtained using time-domain FE simulation and from equations \ref{19} - \ref{22} using $p_1$ and $p_2$ as determined from the analysis (Perturbation \& series, $n=50$) and frequency domain FE simulations (Perturbation \& freq. FE) are compared next. The plots presented in Figs. \ref{fig4a} - \ref{fig4d} show respectively the steady-state axial pressure waveform at time $t=1/(2f_0)$, the mean intensity, amplitude of the pressure component with frequency $\omega=2\pi f_0$, and the amplitude of the pressure component with frequency $2\omega$. Figs. \ref{fig4b} and \ref{fig4c} also show respectively the mean intensity and pressure when $\beta=0$, i.e., for the linear case. Based on the closeness of the responses predicted when $p_1$ and $p_2$ are determined from the analysis and from the FE simulations, it can be concluded that the contribution of NST is not significant. It can also be concluded that the transformation does an excellent job in predicting the pressure distributions at closer distances and that the discrepancies grow as the distance increases. The discrepancy is because the $\epsilon^3-$order secular terms become dominant as the distance increases to the point where  one needs to carry the method of renormalization to $\epsilon^3-$order to maintain the accuracy. Figs. \ref{fig4a} - \ref{fig4d} also show regions where the transformation yields multiple solutions, which invalidates the solution and as such no solution is shown. As noted in previous studies \cite{hamilton1998nonlinear,nayfeh2008nonlinear}, these regions correspond to the shock formation and a shock fitting criteria should be used to obtain the true waveform in these regions. From the results, the transformation yields multiple solutions in a small region around $r/\overline{x}_{5\times10^{-3}}=0.5$, and when $r/\overline{x}_{5\times10^{-3}}>1.13$. However, from Fig. \ref{fig4b}, the mean intensity predicted by the time-domain FE simulation deviates from the linear case only beyond $r/\overline{x}_{5\times10^{-3}}=1.13$. This deviation is a consequence of loss of energy due to the formation and propagation of shock \cite{hamilton1998nonlinear,khokhlova2001numerical}. As such, we hypothesize that $r/\overline{x}_{5\times10^{-3}}=1.13$ is the shock formation location for $\epsilon=5\times10^{-3}$. As expected, an accelerated decrease in $p_{omega}$ is seen after the shock formation at $r/\overline{x}_{5\times10^{-3}}=1.13$ in Fig. \ref{fig4c}. To confirm that a shock didn't occur around $r/\overline{x}_{5\times10^{-3}}=0.5$, we present the time series and Fourier coefficients of the pressure at $r/\overline{x}_{5\times10^{-3}}=0.48$ as obtained from the time-domain FE simulation respectively in Figs. \ref{fig5a} and \ref{fig5b}. The Fourier component of pressure at $2f_0$ is comparable with that at $f_0$ in Fig. \ref{fig5b}, the corresponding time series presented in Fig. \ref{fig5a} doesn't indicate a shock, confirming that no shock occurs in the region around $r/\overline{x}_{5\times10^{-3}}=0.5$. We attribute this anomaly predicted by the model to the local minima in the amplitude of pressure, which is essentially a singularity. As such, multiple solutions are also associated with local minima of the pressure amplitude in the near field. To the knowledge of the authors, this conclusion has not been made in previous investigations as the pressure fields analyzed in those studies did not exhibit local minima.

\begin{figure}[!]
	\begin{center}
		\subfigure[\label{fig6a}] {\includegraphics[width=0.8\textwidth]{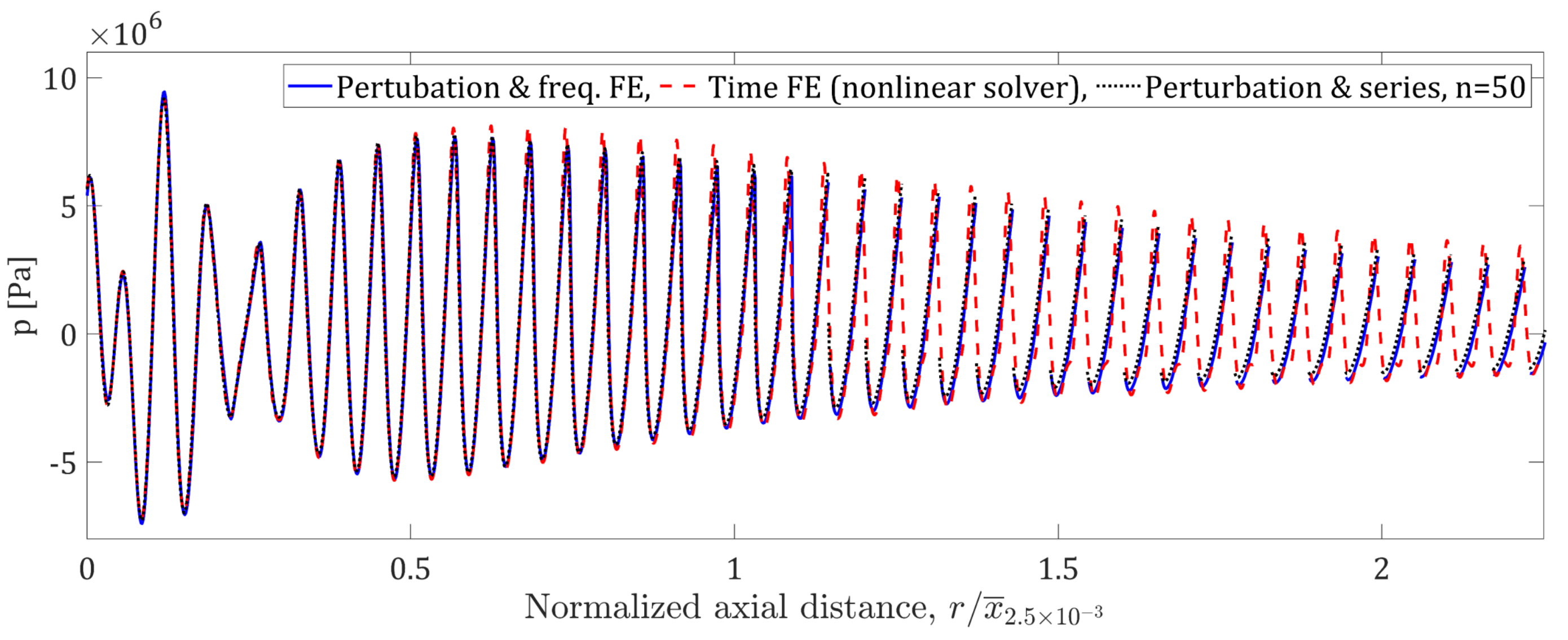}}
		\hspace{0.001in}
		\subfigure[\label{fig6b}] {\includegraphics[width=0.328\textwidth]{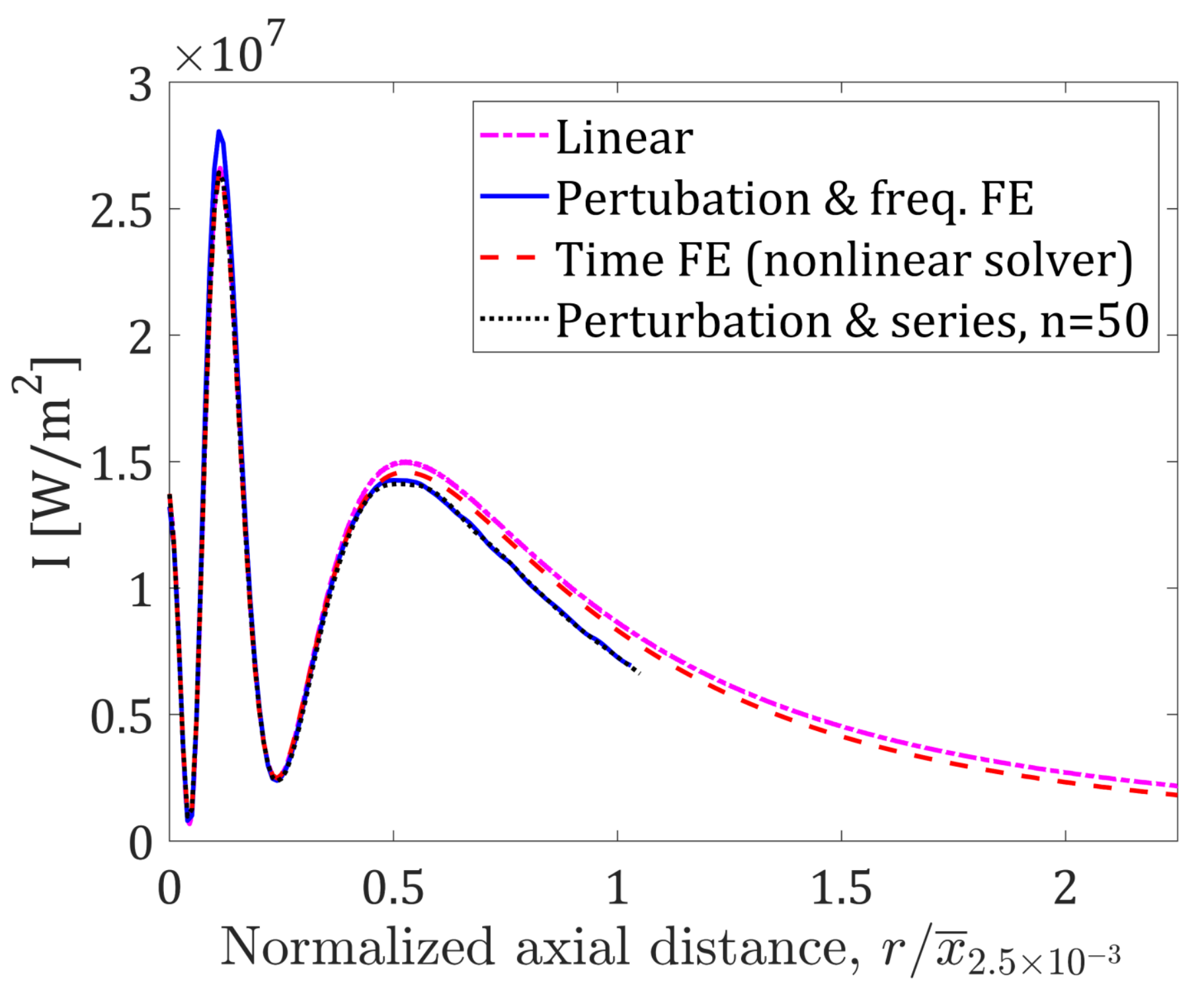}}
		\subfigure[\label{fig6c}] {\includegraphics[width=0.328\textwidth]{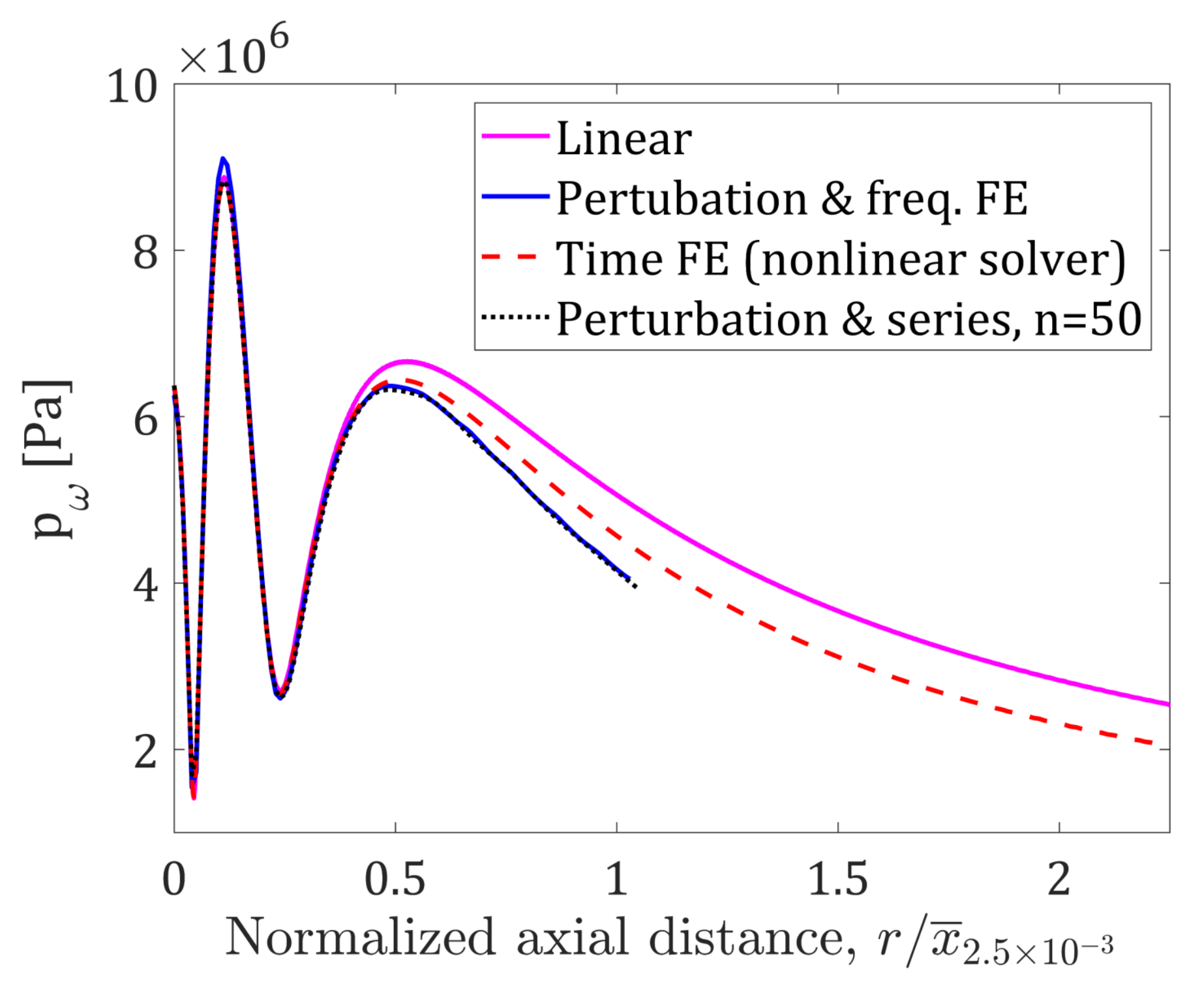}}
		\subfigure[\label{fig6d}] {\includegraphics[width=0.328\textwidth]{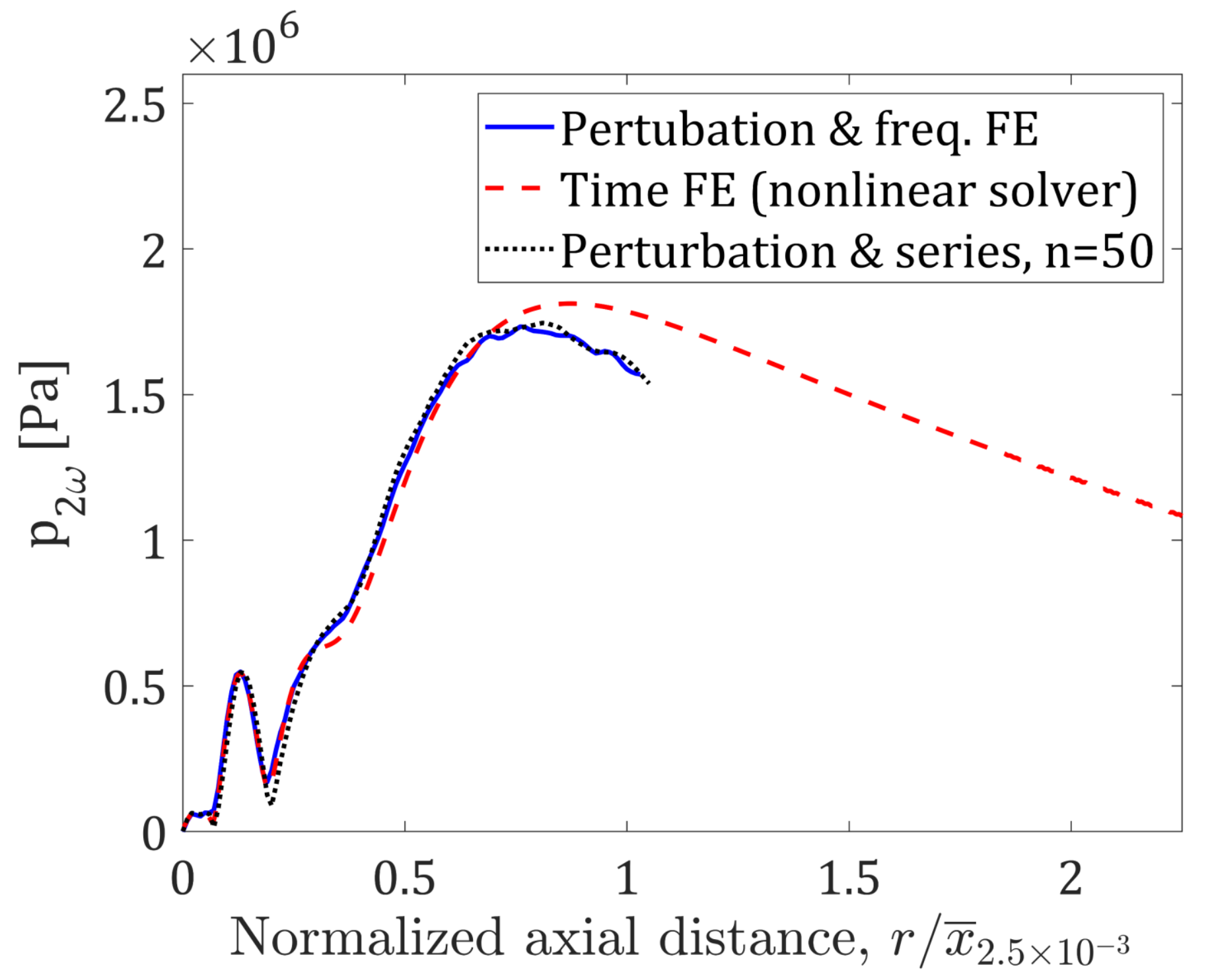}}
	\end{center}
	\vspace{-20pt}
	\caption{Comparison of steady-state axial (a) pressure waveform at time $t=0$, (b) mean intensity, (c) component of the pressure at $\omega=2\pi f_0$, and (d) component of the pressure at $2\omega$ obtained from time-domain FE simulation and from the model developed when $p_1$ and $p_2$ are obtained from the analysis (Perturbation \& series, $n=50$) and from frequency domain simulations (Perturbation \& freq. FE) for $\epsilon=2.5\times10^{-3}$ and excitation frequency $f_0=1.005$ MHz. The axial distance is normalized with the corresponding characteristic shock formation distance $\overline{x}_{2.5\times 10^{-3}}=26.07$ mm.}
\end{figure}

The steady-state axial pressure waveform at time $t=0$, mean intensity, component of pressure at $\omega$, and the component of pressure at $2\omega$ akin to Figs. \ref{fig4a} - \ref{fig4d} for the case of $\epsilon=2.5\times10^{-3}$ are respectively presented in Figs. \ref{fig6a} - \ref{fig6d}. The agreement of the results obtained from the analysis and time-domain FE simulation is similar to that observed in Figs. \ref{fig4a} - \ref{fig4d}. The only major difference is the absence of an anomaly, which suggests that the occurrence of anomaly also depends on the magnitude of $\epsilon$. More importantly, the results predicted by the analysis suggest that the shock formation location on the $z-$axis is $r/\overline{x}_{2.5\times10^{-3}}=1.04$, which is also the point after which the mean intensity predicted by the time-domain FE simulation deviates from the linear case. The characteristic shock formation distance for $\epsilon=2.5\times10^{-3}$ is $\overline{x}_{2.5\times10^{-3}}=26.07$ mm.

\begin{figure}[!]
	\begin{center}
		\subfigure[\label{fig7a}] {\includegraphics[width=0.8\textwidth]{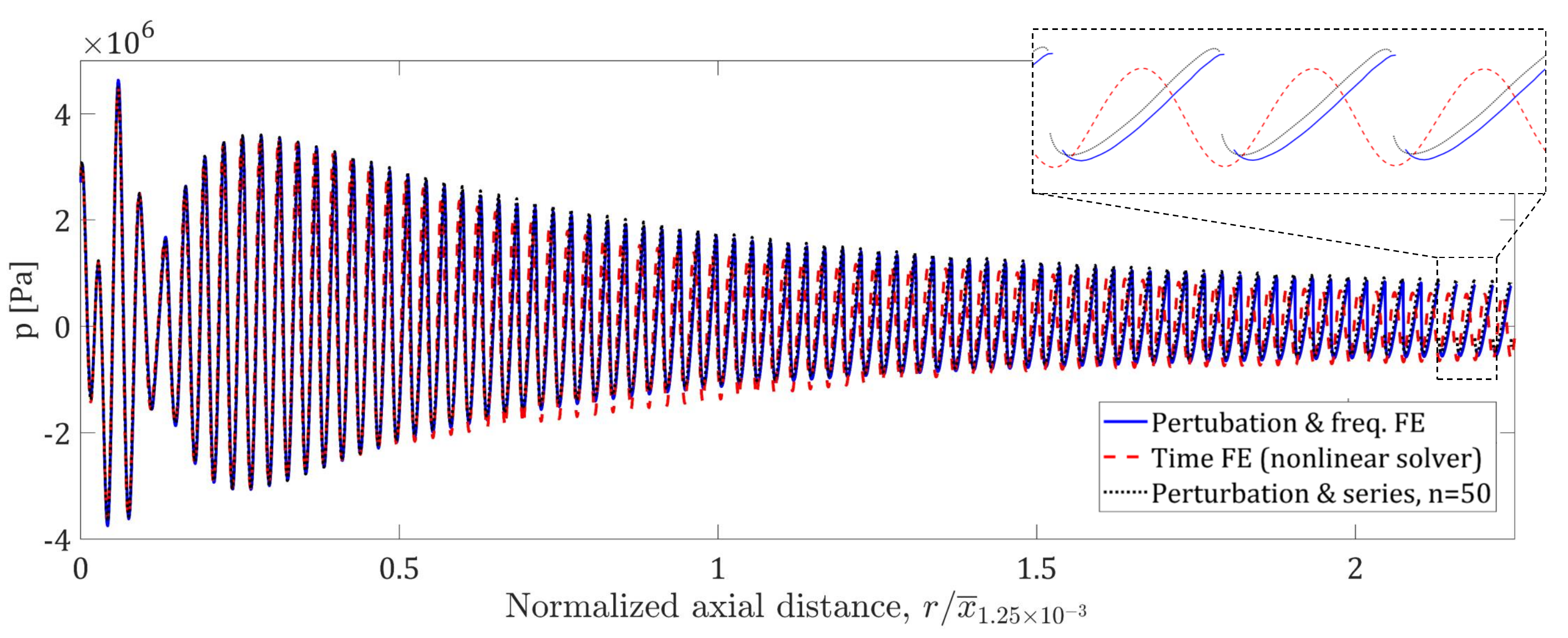}}
		\hspace{0.001in}
		\subfigure[\label{fig7b}] {\includegraphics[width=0.328\textwidth]{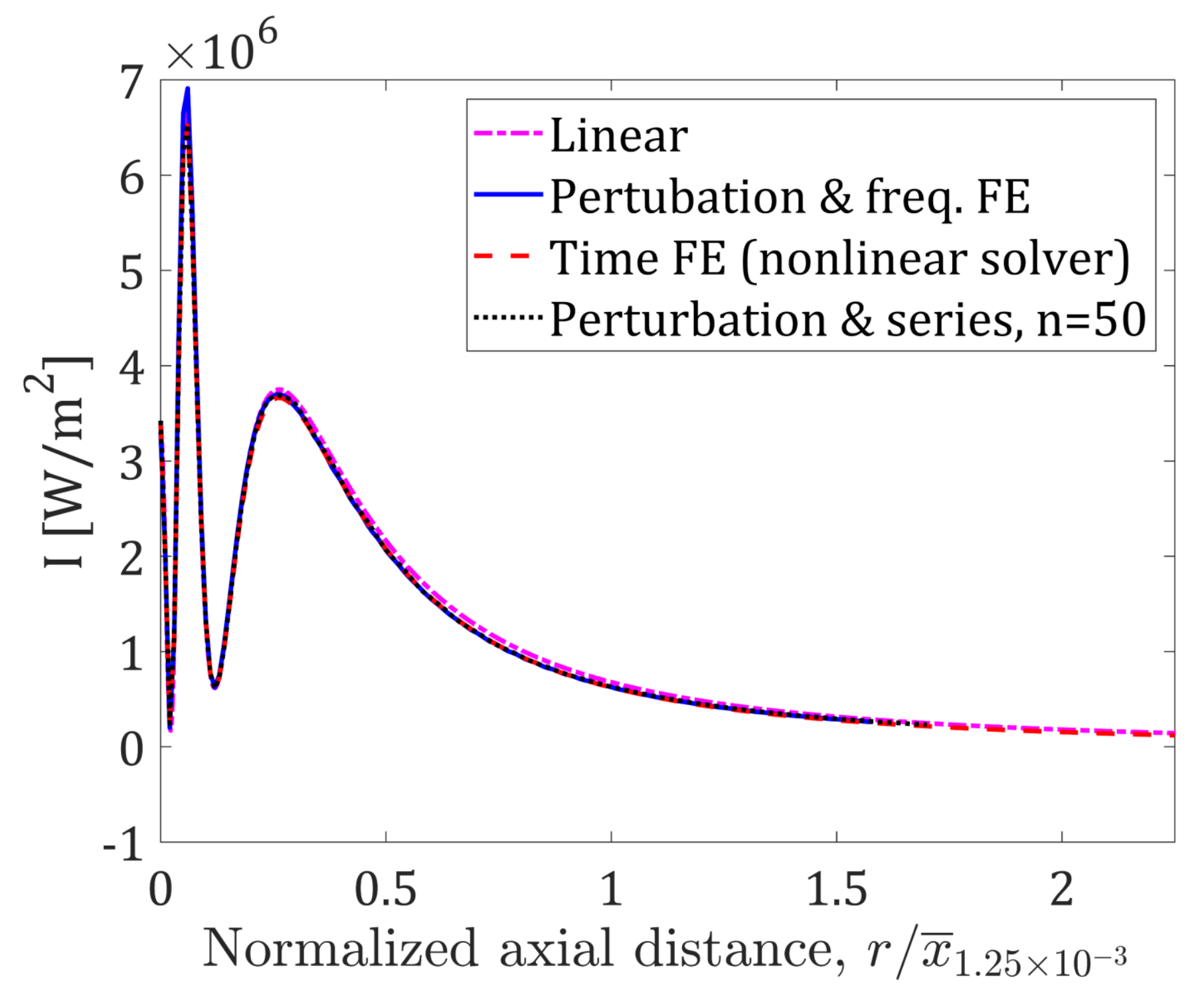}}
		\subfigure[\label{fig7c}] {\includegraphics[width=0.328\textwidth]{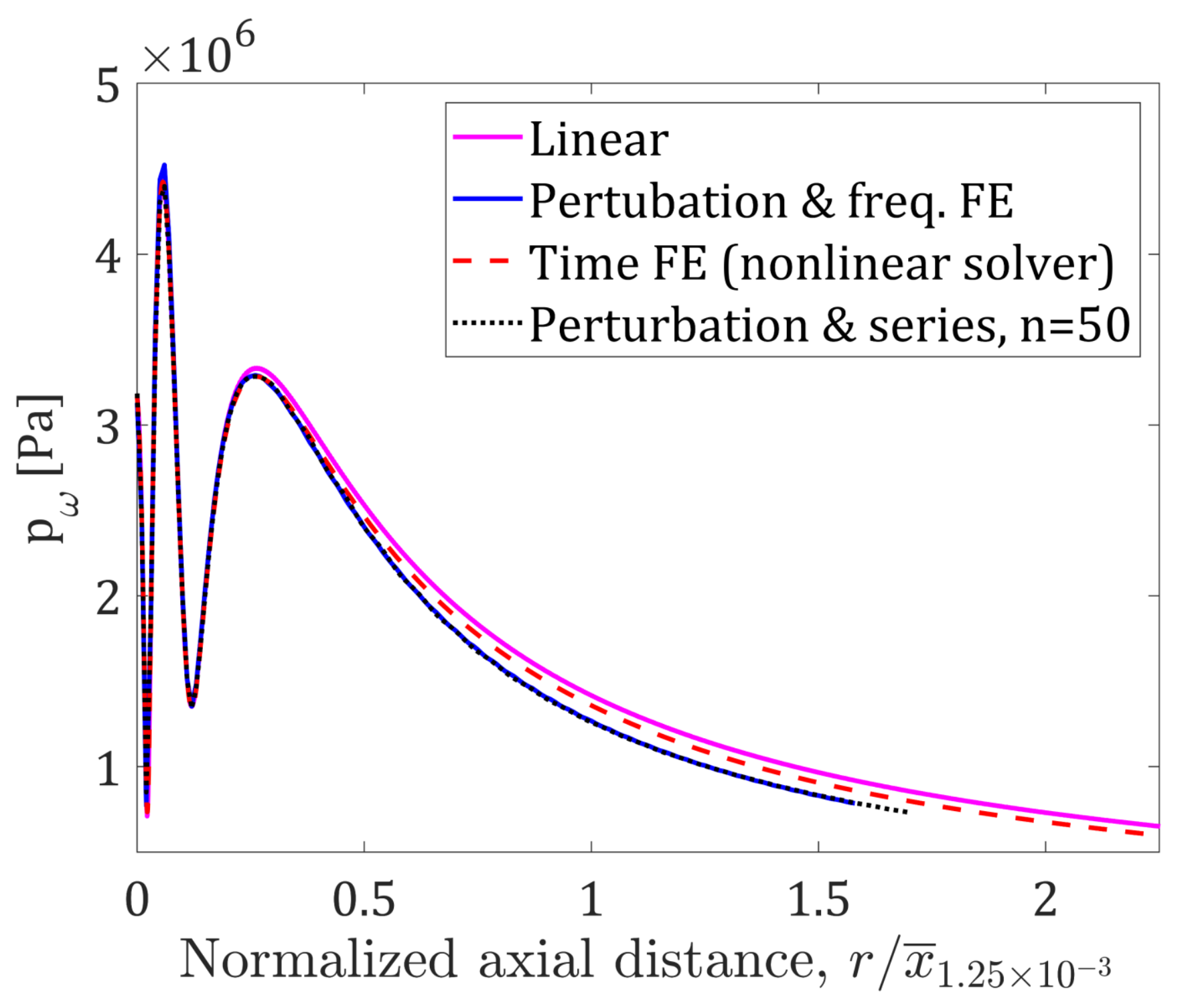}}
		\subfigure[\label{fig7d}] {\includegraphics[width=0.328\textwidth]{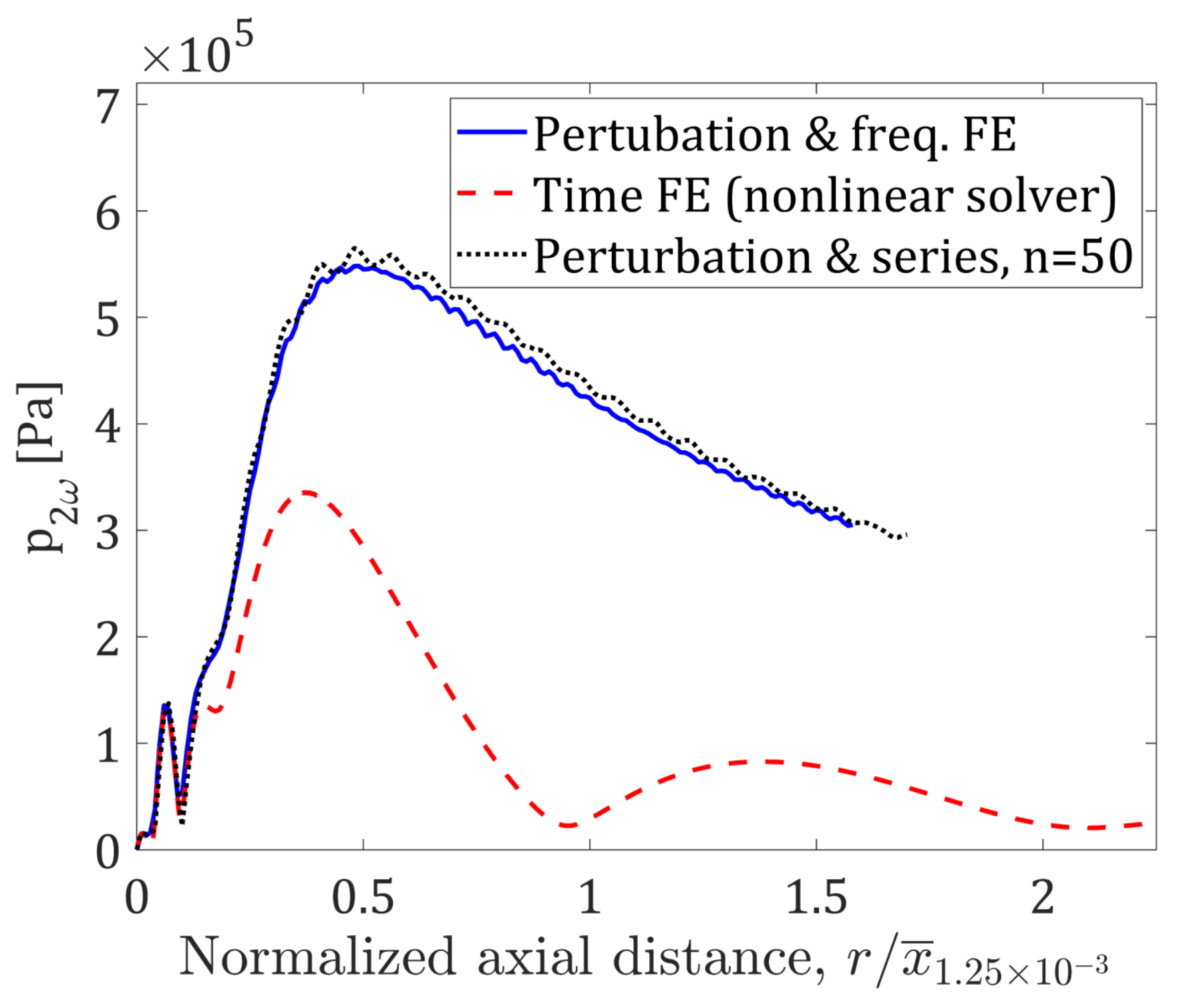}}
	\end{center}
	\vspace{-20pt}
	\caption{Comparison of steady-state axial (a) pressure waveform at time $t=0$, (b) mean intensity, (c) component of the pressure at $\omega=2\pi f_0$, and (d) component of the pressure at $2\omega$ obtained from time-domain FE simulation and from the model developed when $p_1$ and $p_2$ are obtained from analysis (Perturbation \& series, $n=50$) and from frequency domain simulations (Perturbation \& freq. FE) for $\epsilon=1.25\times10^{-3}$ and excitation frequency $f_0=1.005$ MHz. The axial distance is normalized with the corresponding characteristic shock formation distance $\overline{x}_{1.25\times 10^{-3}}=52.13$ mm.}
\end{figure}

The results for the case of excitation amplitude $\epsilon=1.25\times10^{-3}$ are presented in Figs. \ref{fig7a} - \ref{fig7d}. Inspecting the steady-state axial pressure waveform at time $t=0$ predicted by time-domain FE simulation in Fig. \ref{fig7a}, we note that a shock does not take place in the range of $0<r<2.25\times\overline{x}_{1.25\times10^{-3}}$, where $\overline{x}_{1.25\times10^{-3}}=52.13$ mm. This result can also be inferred from Fig. \ref{fig7b}, which doesn't show any deviation of the mean intensity predicted by the time-domain FE simulation from the linear case. Consequently, the deviation of $p_{\omega}$ obtained from time-domain FE simulation from the linear case in Fig. \ref{fig7c} is very small. The results predicted by the analysis suggest that the shock formation location on the $z-$ axis is around $r/\overline{x}_{1.25\times10^{-3}}=1.64$, which falls in the far-field region of the disk. This inconsistency is again a consequence of implementing the method of renormalization only up to the O($\epsilon^2$) order, as discussed earlier.

\subsection{Comparison with previous experiments}
\begin{figure}[!]
	\begin{center}
		\subfigure[\label{fig8a}] {\includegraphics[width=0.4\textwidth]{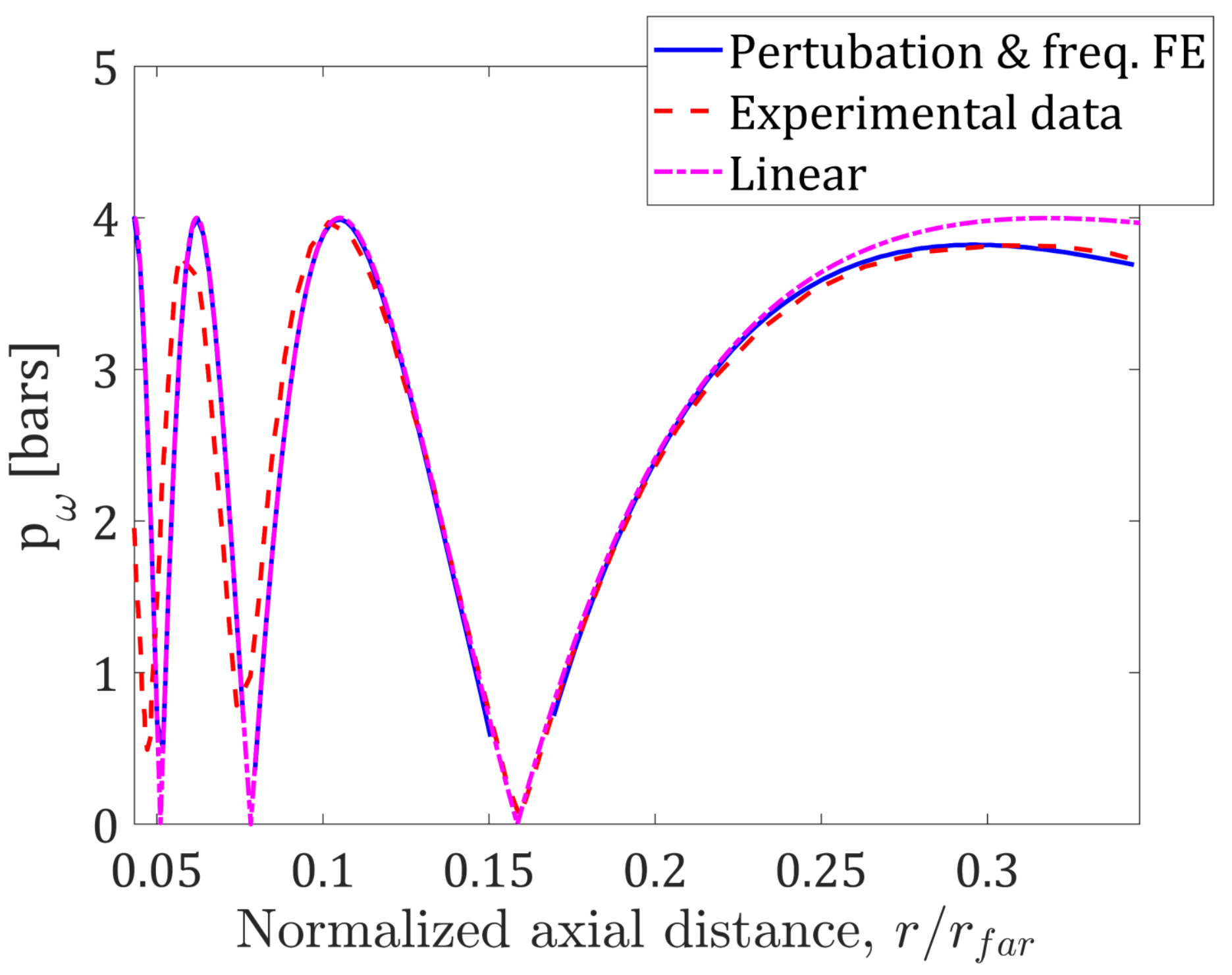}}
		\hspace{0.001in}
		\subfigure[\label{fig8b}] {\includegraphics[width=0.4\textwidth]{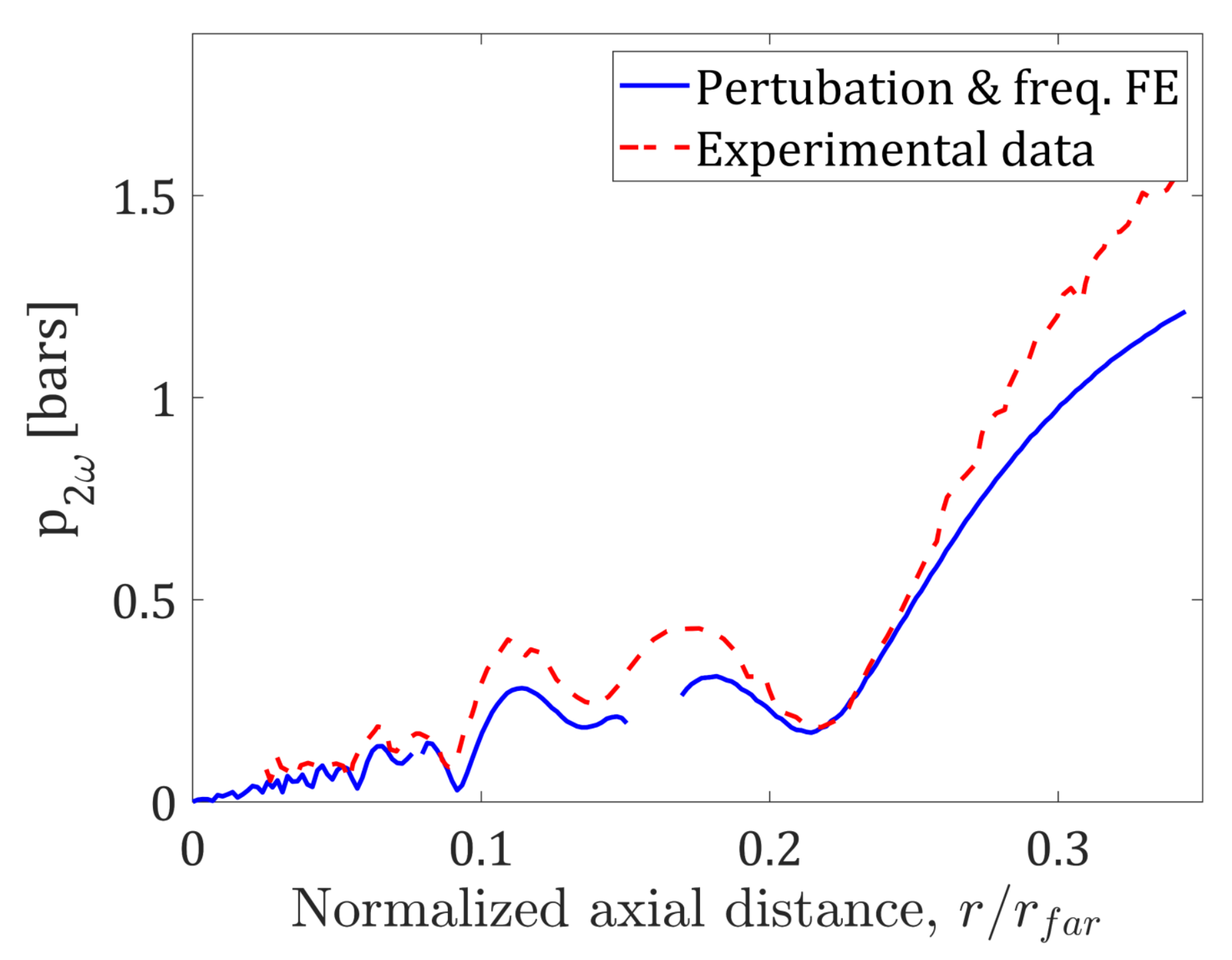}}
		\subfigure[\label{fig8c}] {\includegraphics[width=0.4\textwidth]{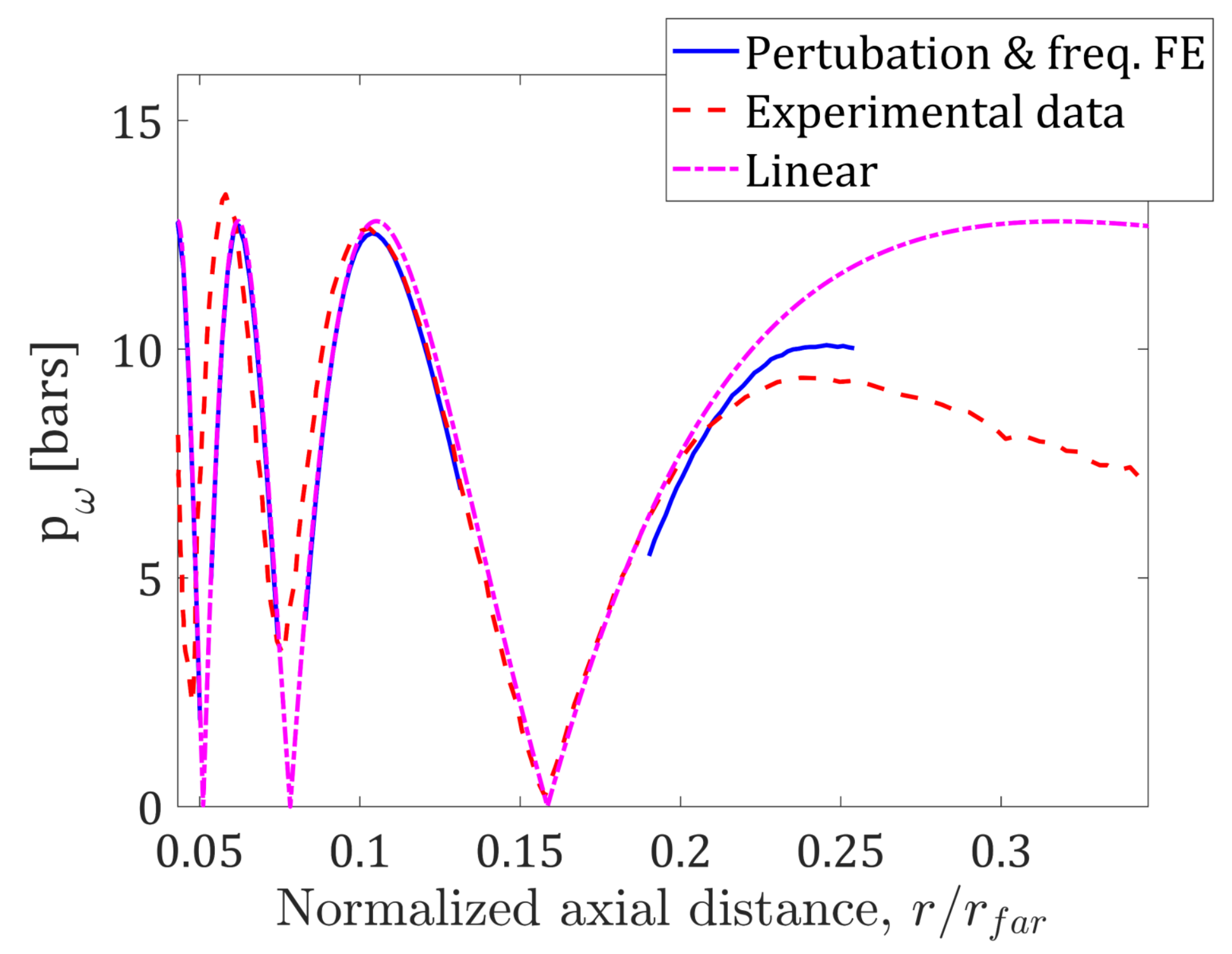}}
		\subfigure[\label{fig8d}] {\includegraphics[width=0.4\textwidth]{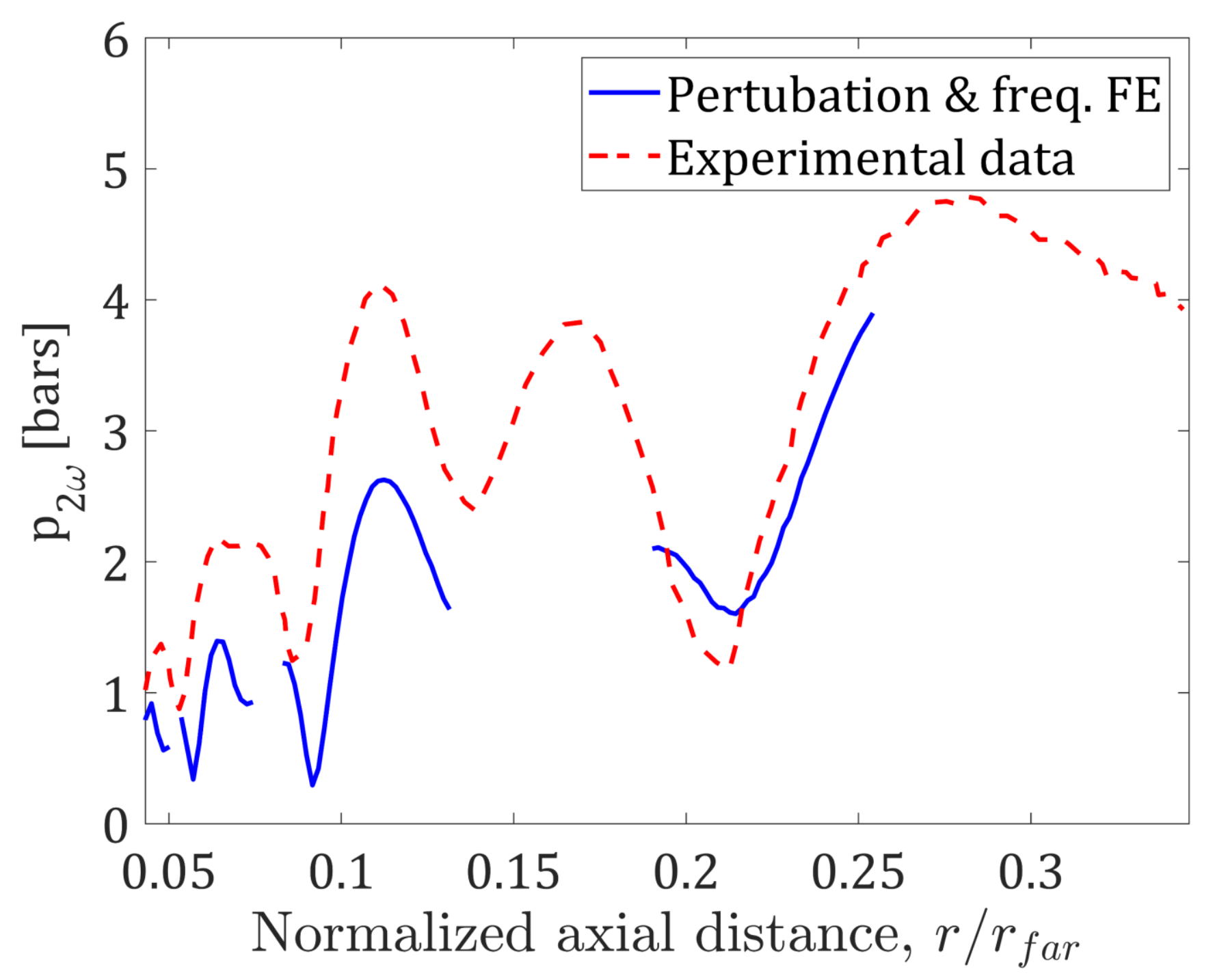}}
		\subfigure[\label{fig8e}] {\includegraphics[width=0.4\textwidth]{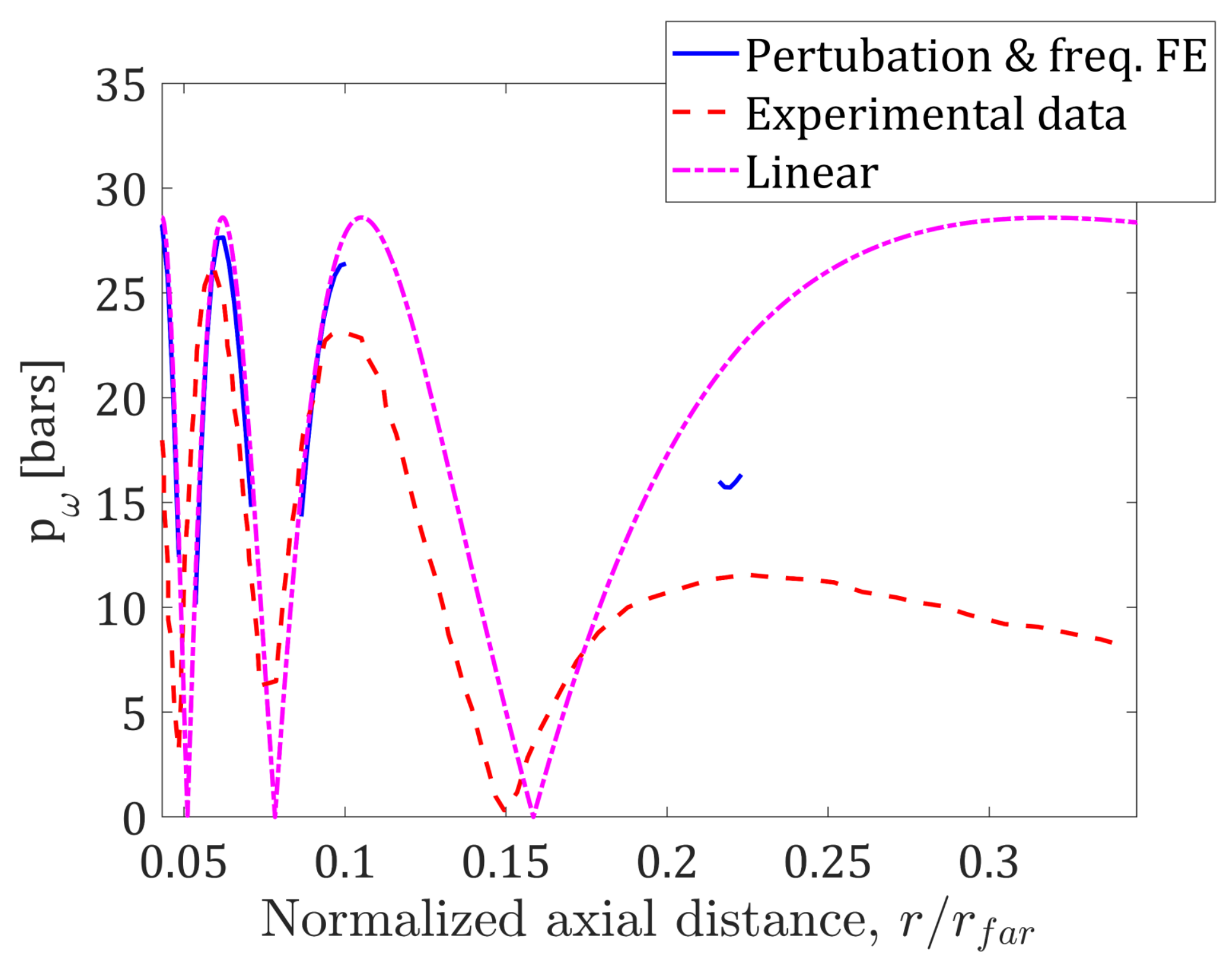}}
		\subfigure[\label{fig8f}] {\includegraphics[width=0.4\textwidth]{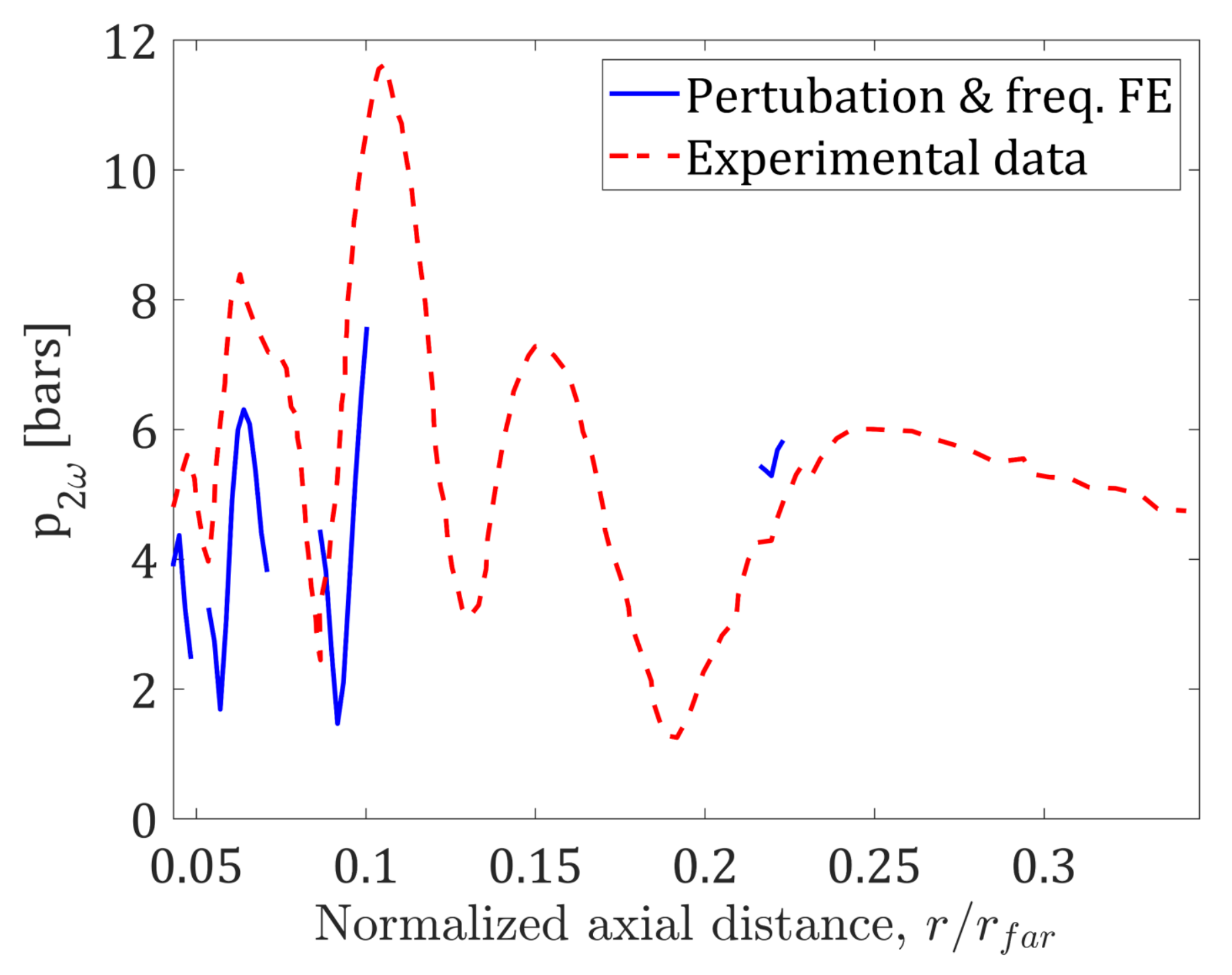}}
	\end{center}
	\vspace{-20pt}
	\caption{Comparisons of Fourier components of the axial pressure generated by a baffled piston at $\omega$ and $2\omega$ predicted using equations \ref{20} - \ref{22} (Perturbation \& freq. FE) with the linear case ($\beta=0$) and corresponding experimental data respectively when (a,b) $p_0=0.2$ MPa, (c,d) $p_0=0.64$ MPa, and (e,f) $p_0=1.43$ MPa. The axial distance is normalized using the Rayleigh far-field distance $r_{\text{far}}=1156.63$ mm and the experimental data is extracted from Khokhlova et al. \cite{khokhlova2001numerical}}
\end{figure}
The efficacy of the developed model in predicting the nonlinear wave propagation and shock formation location is also validated by comparing its predictions with the experimental data of Nachef et al. \cite{nachef1995investigation}. Nachef et al. considered a piezoelectric ceramic disk having a radius of $25$ mm and measured the pressure waveforms over axial distances between $50$ mm and $400$ mm, and lateral distances of between $0$ mm and $50$ mm for different source strengths at an excitation frequency of $f_0=1$ MHz. These waveforms were later treated as a benchmark by Khokhlova et al. \cite{khokhlova2001numerical} to validate their numerical models for $p_0=0.2,0.64,$ and $1.43$ MPa where $p_0=\epsilon\rho_0c_0^2$ is the amplitude of the pressure at the source. They identified that the shock formation distances on the axis are respectively $0.1r_{far}$ and $0.28r_{far}$ for $p_0=1.43$ MPa and $p_0=0.64$ MPa where the Rayleigh far-field distance $r_{far}=1156.63$ mm. They concluded that a shock didn't occur on the axis before $400$ mm \cite{khokhlova2001numerical} when $p_0=0.2$ MPa. The relevant material and geometric properties were identified as $\rho_0=1000$ kg/m$^3$, $c_0=1500$ m/s, $\beta=3.5$, and the equivalent aperture of the circular piston as $a=23.5$ mm \cite{khokhlova2001numerical}. 

From the discussion in Section \ref{sec3p1p2}, $p_1$ and $p_2$ can be evaluated either by the analysis or from appropriate frequency domain linear FE simulations. For the sake of simplicity, we evaluated $p_1$ and $p_2$ using the FE simulations to determine the nonlinear wave propagation generated by a baffled piston. Figs. \ref{fig8a} - \ref{fig8b} show respectively the comparisons of Fourier components of pressure at $\omega$ and $2\omega$  predicted using equations \ref{20} - \ref{22} with the linear case ($\beta=0$) and corresponding experimental data when $p_0=0.2$ MPa. The corresponding results for $p_0=0.64$ MPa and $p_0=1.43$ MPa are presented respectively in Figs. \ref{fig8c} - \ref{fig8d} and Figs. \ref{fig8e} - \ref{fig8f}. From the results, we note the shock formation distances predicted by the method of renormalization are $r/r_{far}=0.25$, and $r/r_{far}=0.1$ respectively for $p_0=0.64$ MPa and $p_0=1.43$ MPa and that a shock did not occur in the range considered for $p_0=0.2$ MPa. These findings are are very close to the shock formation distances identified by Khokhlova et al. \cite{khokhlova2001numerical}, which validate the analysis.

%%%%%%%%%%%%%%%%%%%%%%%%%%%%%%%%%%%%%%%%%%%%%%%%%%%%%%%%%%%%%%%%%%%%%%%%%%%%%%%%%%%%%%%%%%%%%%%%%%%%%%%%%%%%%%%%%%%%%%%%%%%%%%%%%%%%%%%%%%%%%%%%%%%%%%%%%%%%%%%%%%%%%%%%%%%%%%%%%%%%%%%%%%%%%%%%%%%%%%%%%%%%%%%%%%%%%%%%%%%%%%%%%%%%%%%%%%%%%%%%%%%%%%%%%%%%%%%%%%%%%%%%%%%%%%%%%%%%%%%%%%%%%%%%%%%%%%%%%%%%%%%%%%%%%%%%%%%%%%%%%%%%%%%%%%%%%%%%

\section{Conclusions}
The knowledge of the shock formation distance is very crucial for AET systems and the analysis developed in this work provides a quick estimate that can aid the design and operation of efficient AET systems. The analysis is based on asymptotic expansion of the governing equation. The method of renormalization  is used to predict the nonlinear wave generated by a finite amplitude baffled circular disk for a specified deformation profile. The lossless form of the Westervelt equation and the normal velocity continuity boundary condition were scaled with a non-dimensional parameter related to the amplitude of the transverse displacement of the disk. They were expanded and solved by introducing a coordinate transformation to eliminate the secular terms, which resulted in singularities that required redefining the relation between the pressure components. We validated the analysis by comparing its predictions of the shock formation distance under different excitation levels with higher fidelity nonlinear finite element simulations and previously published experimental results. We also demonstrated the versatility of the analysis by showing that the $\epsilon-$order and $\epsilon^2-$order solutions can be determined either from analytical expressions or from linear frequency domain finite element simulations. We showed that the analysis accurately predicts the nonlinear wave propagation and shock formation distance in the near-field of the disk and argued that the accuracy level can be increased by extending the analysis to $\epsilon^3$ or higher orders. We found out that the transformation yields multiple solutions at the shock location and at local minima that occur in the near-field of the pressure. We also showed an accelerated decrease in the power of the excitation frequency beyond the shock formation. Of particular importance is the validity of the solution for any excitation level because the $\epsilon$ and $\epsilon^2$ order solutions are independent of that level. 

%%%%%%%%%%%%%%%%%%%%%%%%%%%%%%%%%%%%%%%%%%%%%%%%%%%%%%%%%%%%%%%%%%%%%%%%%%%%%%%%%%%%%%%%%%%%%%%%%%%%%%%%%%%%%%%%%%%%%%%%%%%%%%%%%%%%%%%%%%%%%%%%%%%%%%%%%%%%%%%%%%%%%%%%%%%%%%%%%%%%%%%%%%%%%%%%%%%%%%%%%%%%%%%%%%%%%%%%%%%%%%%%%%%%%%%%%%%%%%%%%%%%%%%%%%%%%%%%%%%%%%%%%%%%%%%%%%%%%%%%%%%%%%%%%%%%%%%%%%%%%%%%%%%%%%%%%%%%%%%%%%%%%%%%%%%%%%%%
\section{Compliance with Ethical Standards - Conflict of interest}
The authors declare that they have no conflict of interest.
%%%%%%%%%%%%%%%%%%%%%%%%%%%%%%%%%%%%%%%%%%%%%%%%%%%%%%%%%%%%%%%%%%%%%%%%%%%%%%%%%%%%%%%%%%%%%%%%%%%%%%%%%%%%%%%%%%%%%%%%%%%%%%%%%%%%%%%%%%%%%%%%%%%%%%%%%%%%%%%%%%%%%%%%%%%%%%%%%%%%%%%%%%%%%%%%%%%%%%%%%%%%%%%%%%%%%%%%%%%%%%%%%%%%%%%%%%%%%%%%%%%%%%%%%%%%%%%%%%%%%%%%%%%%%%%%%%%%%%%%%%%%%%%%%%%%%%%%%%%%%%%%%%%%%%%%%%%%%%%%%%%%%%%%%%%%%%%%
\section{Acknowledgments}
This work was supported by the National Science Foundation Grant No. ECCS$-1711139$, which is gratefully acknowledged. The authors would also like to thank Professor Saad Ragab (Virginia Tech) for providing valuable feedback in developing the model. 
%%%%%%%%%%%%%%%%%%%%%%%%%%%%%%%%%%%%%%%%%%%%%%%%%%%%%%%%%%%%%%%%%%%%%%%%%%%%%%%%%%%%%%%%%%%%%%%%%%%%%%%%%%%%%%%%%%%%%%%%%%%%%%%%%%%%%%%%%%%%%%%%%%%%%%%%%%%%%%%%%%%%%%%%%%%%%%%%%%%%%%%%%%%%%%%%%%%%%%%%%%%%%%%%%%%%%%%%%%%%%%%%%%%%%%%%%%%%%%%%%%%%%%%%%%%%%%%%%%%%%%%%%%%%%%%%%%%%%%%%%%%%%%%%%%%%%%%%%%%%%%%%%%%%%%%%%%%%%%%%%%%%%%%%%%%%%%%%

\appendix
\section{$\epsilon^2-$order solution} \label{appa}
To determine the particular solution of equation \ref{13}, we use the separation of variables technique and assume the solution as 
\begin{equation}
p_2(r_s,\theta_s,t)=
2\sum_{n=0}^{\infty}\sum_{m=0}^{\infty}\sum_{q=0}^{n+m}(2n+1)(2m+1)k^4\beta\rho_0\omega^2\kappa_{qnm}P_q(\cos\theta_s) \left[\frac{ }{ }\sin \left(2 \omega t \right)\Psi_{qnm}(rr_s)+\cos \left(2 \omega t \right)X_{qnm}(r_s)\frac{ }{ }\right]
\label{A1}
\end{equation}
Substitution of equation \ref{A1} in equation \ref{13} yields
\begin{subequations}
	\begin{equation}
\left[\frac{1}{r^2}q(q+1)\Psi_{qnm}(r_s) -\frac{2}{r}\frac{\partial \Psi_{qnm}(r_s)}{\partial r_s} -\frac{\partial^2 \Psi_{qnm}(r_s)}{\partial r_s^2}\right]
		-4k^2\Psi_{qnm}(r_s)=-E_{\Psi}(r_s)
	\label{A2a}
	\end{equation}
	\begin{equation}
	\left[\frac{1}{r^2}q(q+1)X_{qnm}(r_s) -\frac{2}{r}\frac{\partial X_{qnm}(r_s)}{\partial r_s} -\frac{\partial^2 X_{qnm}(r_s)}{\partial r_s^2}\right]
	-4k^2X_{qnm}(r_s)=-E_X(r_s)
	\label{A2b}
	\end{equation}
	where
	\begin{multline}
		E_\Psi(r_s)=\left(\Delta_n^{(c)}\Delta_m^{(c)}-\Delta_n^{(s)}\Delta_m^{(s)}\right)\left(\frac{ }{ }j_m(kr_s)y_n(kr_s)+j_n(kr_s)y_m(kr_s)\frac{ }{ }\right)\\
		-\left(\Delta_n^{(c)}\Delta_m^{(s)}+\Delta_n^{(s)}\Delta_m^{(c)}\right)\left(\frac{ }{ }j_n(kr_s)j_m(kr_s)-y_n(kr_s)y_m(kr_s)\frac{ }{ }\right)
	\end{multline}
	\begin{multline}
		E_X(r_s)=\left(\Delta_n^{(c)}\Delta_m^{(c)}-\Delta_n^{(s)}\Delta_m^{(s)}\right)\left(\frac{ }{ }j_n(kr_s)j_m(kr_s)
		-y_n(kr_s)y_m(kr_s)\frac{ }{ }\right)\\
		+\left(\Delta_n^{(c)}\Delta_m^{(s)}+\Delta_n^{(s)}\Delta_m^{(c)}\right)\left(\frac{ }{ }j_m(kr_s)y_n(kr_s)+j_n(kr_s)y_m(kr_s)\frac{ }{ }\right)
\end{multline}
\end{subequations}
The homogeneous solutions of equations \ref{A2a} and \ref{A2b} are $X_1(r_s)=\Psi_1(r_s)=j_q(2kr_s)$ and $X_2(r_s)=\Psi_2(r_s)=y_q(2kr_s)$, which are now used to determine its particular solution by using the variation of parameters technique as
\begin{subequations}
	\begin{multline}
	\Psi_{qnm}(r_s)=
	\Psi_1(r_s)\int\displaylimits_{r_0/\cos\theta_s}^{r_s}\frac{\Psi_2(x)}{\Psi_1(x)\frac{\partial \Psi_2(x) }{\partial x}-\Psi_2(x)\frac{\partial \Psi_1(x) }{\partial x}}E_{\Psi}(x)\,dx\\
	-\Psi_2(r_s)\int\displaylimits_{r_0/\cos\theta_s}^{r_s}\frac{\Psi_1(x)}{\Psi_1(x)\frac{\partial \Psi_2(x) }{\partial x}-\Psi_2(x)\frac{\partial \Psi_1(x) }{\partial x}}E_{\Psi}(x)\,dx
	\label{A3a}
	\end{multline}
	\begin{multline}
	X_{qnm}(r_s)=X_1(r_s)\int\displaylimits_{r_0/\cos\theta_s}^{r_s}\frac{X_2(x)}{X_1(x)\frac{\partial X_2(x) }{\partial x}-X_2(x)\frac{\partial X_1(x) }{\partial x}}E_{X}(x)\,dx\\
	-X_2(r_s)\int\displaylimits_{r_0/\cos\theta_s}^{r_s}\frac{X_1(x)}{X_1(x)\frac{\partial X_2(x) }{\partial x}-X_2(x)\frac{\partial X_1(x) }{\partial x}}E_{X}(x)\,dx
	\label{A3b}
	\end{multline}
	where $x$ is a dummy variable.
\end{subequations}
Since the closed form expression of spherical Bessel functions in terms of trigonometric functions are required to identify the secular terms in $\epsilon^2-$order solution (i.e., $\Psi_{qnm}(r_s)$ and $X_{qnm}(r_s)$), we use an identity of Rayleigh's formula and represent spherical Bessel function of the first kind as
\begin{equation}
    j_n(x) =
    \sum_{n_1=0}^{n}\begin{bmatrix}n\\n_1\end{bmatrix}2^{n-n_1}\sum_{n_2=0}^{n_1}n_2!\begin{Bmatrix}n_1\\n_2\end{Bmatrix}\sum_{n_3=0}^{n_2}\frac{(-1)^{n+n_2-n_3}x^{-n-1-n_3}\sin(x+n_3\pi/2)}{n_3!}
    \label{A4a}
\end{equation}
where $\begin{bmatrix}n\\n_1\end{bmatrix}$ and $\begin{Bmatrix}n_1\\n_2\end{Bmatrix}$ are respectively the Stirling numbers of the first and second kind. From equation \ref{A4a}, the coefficients of the trigonometric functions can be determined which can then be used to determine $y_n(x)$. By using the so obtained analytical expressions of the spherical Bessel functions, it can be showed that
\begin{equation}
		\left[\Psi_1(x)\frac{\partial \Psi_2(x) }{\partial x}-\Psi_2(x)\frac{\partial \Psi_1(x) }{\partial x}\right]=\left[X_1(x)\frac{\partial X_2(x) }{\partial x}-X_2(x)\frac{\partial X_1(x) }{\partial x}\right]=\frac{1}{2kx^2}
\end{equation}
Using the above relation, equations \ref{A3a} and \ref{A3b} are rewritten as
\begin{subequations}
 	\begin{equation}
	\Psi_{qnm}(r_s)=	\Psi_1(r_s)\int\displaylimits_{r_0/\cos\theta_s}^{r_s}2kx^2\Psi_2(x)E_{\Psi}(x)\,dx
	-\Psi_2(r_s)\int\displaylimits_{r_0/\cos\theta_s}^{r_s}2kx^2\Psi_1(x)E_{\Psi}(x)\,dx
	\label{A6a}
	\end{equation}
	\begin{equation}
	X_{qnm}(r_s)=	X_1(r_s)\int\displaylimits_{r_0/\cos\theta_s}^{r_s}2kx^2X_2(x)E_{X}(x)\,dx
	-\Psi_2(r_s)\int\displaylimits_{r_0/\cos\theta_s}^{r_s}2kx^2X_1(x)E_{X}(x)\,dx
	\label{A6b}
	\end{equation} 
\end{subequations}
Evaluation of integrals in equations \ref{A6a} and \ref{A6b} shows that $\Psi_{qnm}(r_s)$ and $X_{qnm}(r_s)$ contain sine-integral, cosine-integral, $1/r_s^{[\dots]}$, and logarithmic terms. Furthermore, it can also be shown that logarithmic terms grow indefinitely for large $r_s$ with respect to $\epsilon-$order solution and are the secular terms. Although the rest of the terms do not grow indefinitely for large $r$, some of them have a major contribution to the $\epsilon^2-$order response, especially at closer distances to the disk, and hence govern the nonlinear interaction. As such, we note that an asymptotic form (such as Kelly and Nayfeh \cite{kelly1980non}) of these expressions will not convey accurate physics. Furthermore, we noticed that sine-integral and cosine-integral terms decay monotonically at a faster rate than logarithmic and $1/r_s^{[\dots]}$ terms and hence we regard them as non-secular terms (NST) and are neglected in the further analysis.

It is relevant to point out that the homogeneous solution of $\epsilon^2-$order equation that is responsible for satisfying the $\epsilon^2-$order boundary condition is not considered in the analysis as it represents a freely propagating wave arising from excitation at the boundary and hence doesn't exhibit an unbounded growth (i.e., NST). Moreover, due to equations \ref{A3a} and \ref{A3b}, the $\epsilon^2-$order vanishes on the surface of the disk ($r_s=r_0/\cos\theta_s$). This condition ensures that any transformation defined in the method of renormalization vanishes on the surface of the disk, as is required to describe accurate underlying physics \cite{kelly1980non}. Furthermore, from the definitions of $\Delta^{(c)}_n$ and $\Delta^{(c)}_n$, we note that they are the only parameters that vary for a different deformation profile and that they are constants of integration in determining $\Psi_{qnm}(r_s)$ and $X_{qnm}(r_s)$. As such, the analytical expressions for these integrals can be utilized to determine $p_2$ for any deformation pattern.

%\bibliographystyle{acm}
%\bibliography{vc24vt}

\end{document}